\documentclass[fleqn,usenatbib]{mnras}
\usepackage[utf8]{inputenc}
\usepackage[T1]{fontenc}

\usepackage{newtxtext,newtxmath}
\DeclareRobustCommand{\VAN}[3]{#2}
\let\VANthebibliography\thebibliography
\def\thebibliography{\DeclareRobustCommand{\VAN}[3]{##3}\VANthebibliography}

\usepackage{amsmath,cleveref,flushend,hhline}
\usepackage{siunitx}
\DeclareSIUnit{\au}{au}
\DeclareSIUnit{\year}{yr}

\usepackage[dvipsnames]{xcolor}
\usepackage{tikz}
\usepackage{3dplot}

\hypersetup{pdfauthor={B. V. Lehmann et al.},
    pdftitle={Three-body capture, ejection, and the demographics of bound objects in binary systems},
    pdfkeywords={},
    bookmarksnumbered=true,
    linkcolor=blue,
    citecolor=blue,
    filecolor=cyan,
    urlcolor=Mulberry
}

\usepackage{bm}
\newcommand{\bb}[1]{\bm{\mathrm{#1}}}

\newcommand{\du}{\mathrm{d}}
\newcommand{\dd}{\,\du}
\newcommand{\arctanh}{\operatorname{arctanh}}
\newcommand{\sign}{\operatorname{sign}}

\newcommand\rclose{r_{\mathrm{close}}}
\newcommand\vesc{v_{\mathrm{esc}}}

\newcommand\scap{\sigma_{\mathrm{cap}}}
\newcommand\sej{\sigma_{\mathrm{ej}}}
\newcommand\Rej{R_{\mathrm{ej}}}
\newcommand\Rcap{R_{\mathrm{cap}}}

\newcommand\orcid[1]{\href{http://orcid.org/#1}
    {\hspace{2pt}\ignorespaces
        \textsuperscript{\includegraphics[width=8pt]{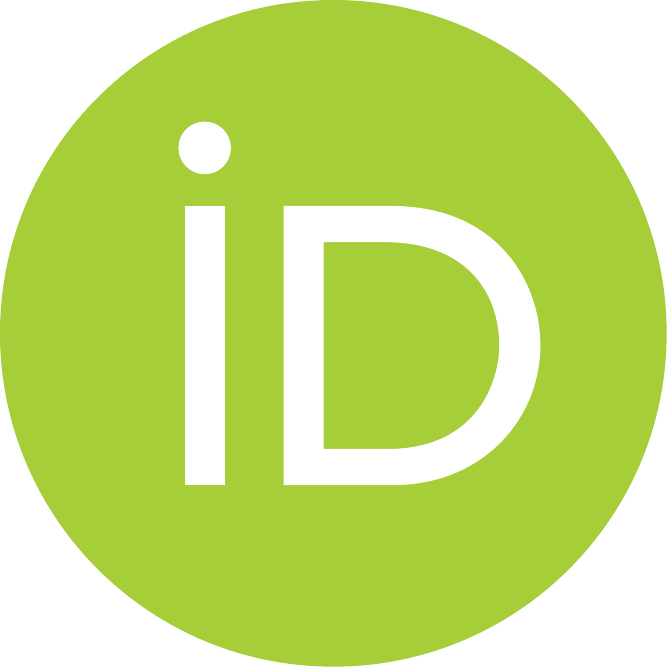}}}}
\newcommand\email[1]{\href{mailto:#1}{\textcolor{blue}{#1}}}

\title[Three-body capture and bound objects]{Three-body capture, ejection, and the demographics of bound objects in binary systems}

\author[B. V. Lehmann et al.]{
Benjamin V. Lehmann\ignorespaces
    \orcid{0000-0001-7735-4673},\ignorespaces
    \thanks{E-mail: \email{benvlehmann@gmail.com}}
Olivia G. Ross\ignorespaces
    \orcid{0000-0001-6778-6659},\ignorespaces
    \textsuperscript{\thanks{E-mail: \email{oross@ucsc.edu}}}
Ava Webber\ignorespaces
    \orcid{0000-0003-1553-2280}\ignorespaces
    \textsuperscript{\thanks{E-mail: \email{arwebber@ucsc.edu}}}
and Stefano Profumo\ignorespaces
    \orcid{0000-0002-9159-7556}\ignorespaces
    \textsuperscript{\thanks{E-mail: \email{profumo@ucsc.edu}}}
\\
Department of Physics, University of California Santa Cruz, 1156 High St, Santa Cruz, CA 95064, USA
\\
Santa Cruz Institute for Particle Physics, 1156 High St, Santa Cruz, CA 95064, USA}

\date{Accepted XXX. Received YYY; in original form ZZZ}
\pubyear{2021}

\begin{document}
\label{firstpage}
\pagerange{\pageref{firstpage}--\pageref{lastpage}}
\maketitle

\begin{abstract}
    We study the capture of light objects of arbitrary velocity by binary systems. Extending results for the capture of comets in the solar system, we develop a simple geometric characterization of the capture cross section, leading directly to the distribution of orbital parameters of captured objects. We use the same framework to study the lifetimes of these bound orbits prior to ejection, and find that a simplified version of the \"Opik--Arnold approach readily yields a closed-form estimate for the ejection rate that agrees well with numerical experiments. Without any detailed-balance assumptions, our results make manifest the characteristics of close encounters leading to capture and ejection. As an application of our results, we demonstrate the estimation of the equilibrium population of captured dark matter particles in a binary system.
\end{abstract}


\section{Introduction}
Recent observational advances have led to significant developments in our understanding of extrasolar binary systems. Among the most prominent of these are the numerous recently discovered extrasolar planetary systems \cite[see e.g.][]{2007ARA&A..45..397U,Schneider:2011vr,Wright:2010bc,Cassan:2012jx,Akeson:2013mqa,2018ApJS..235...38T}. But there are other notable examples with direct relevance to fundamental physics: precision studies of the dynamics of pulsar binary systems probe a variety of new physics scenarios through their timing signatures \cite[see e.g.][]{Stairs:2003eg,Cordes:2005gp,Hobbs:2009yy,Hobbs:2017oam,Cordes:2019lok,Siemens:2019xkk}, and the gravitational waves sourced by compact object binaries provide an entirely new observable for astrophysics, cosmology, and particle physics (see e.g. \citet{Sopuerta:2014hra,Baiotti:2016qnr} for recent reviews). For each class of binaries, it is essential to understand the mechanisms by which such systems evolve and interact with their environments.

\begin{figure}\centering
    \includegraphics[width=0.45\columnwidth, trim=1cm 1.2cm 3cm 0, clip]{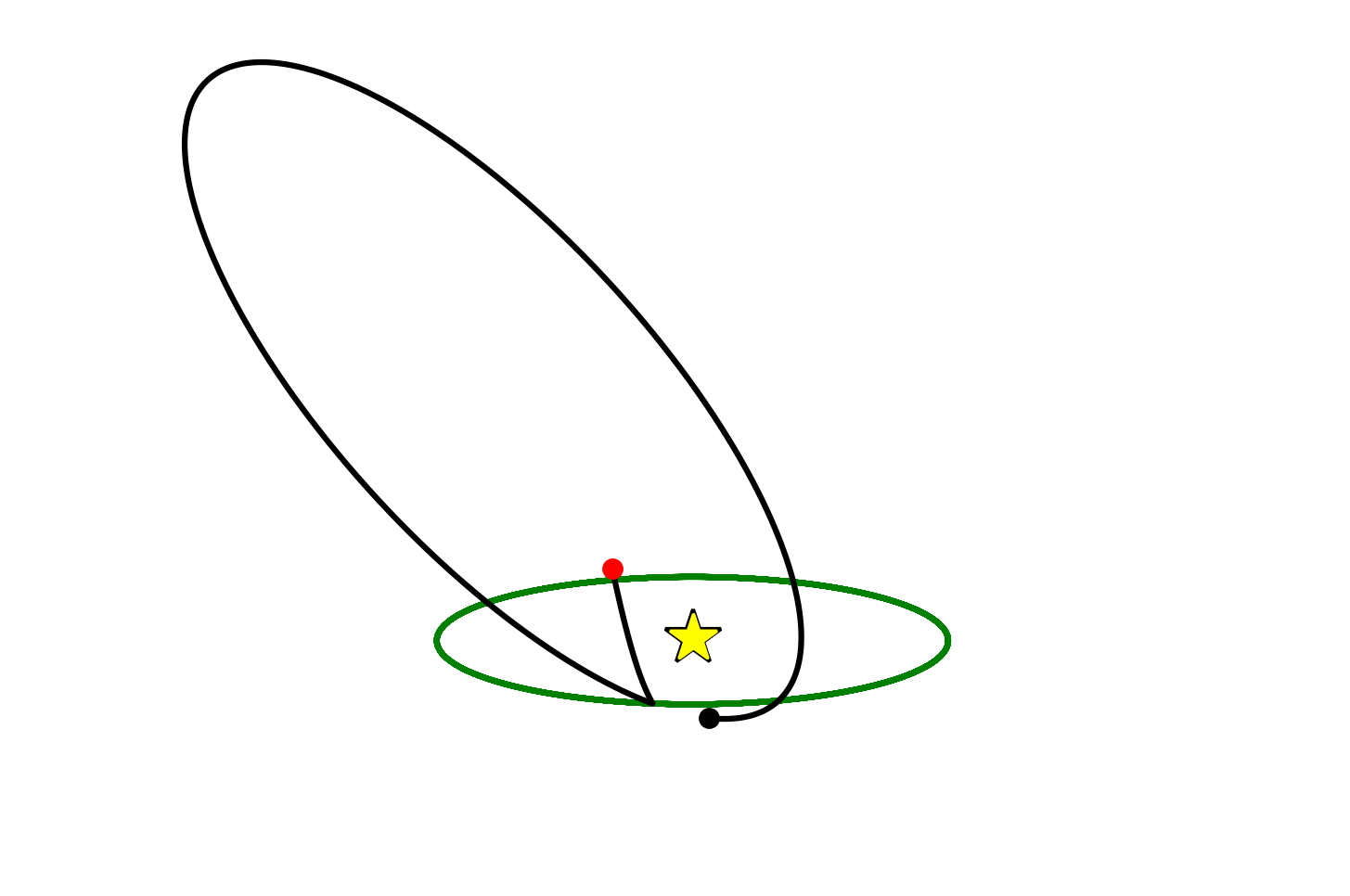}
    \includegraphics[width=0.45\columnwidth, trim=1cm 1.2cm 3cm 0, clip]{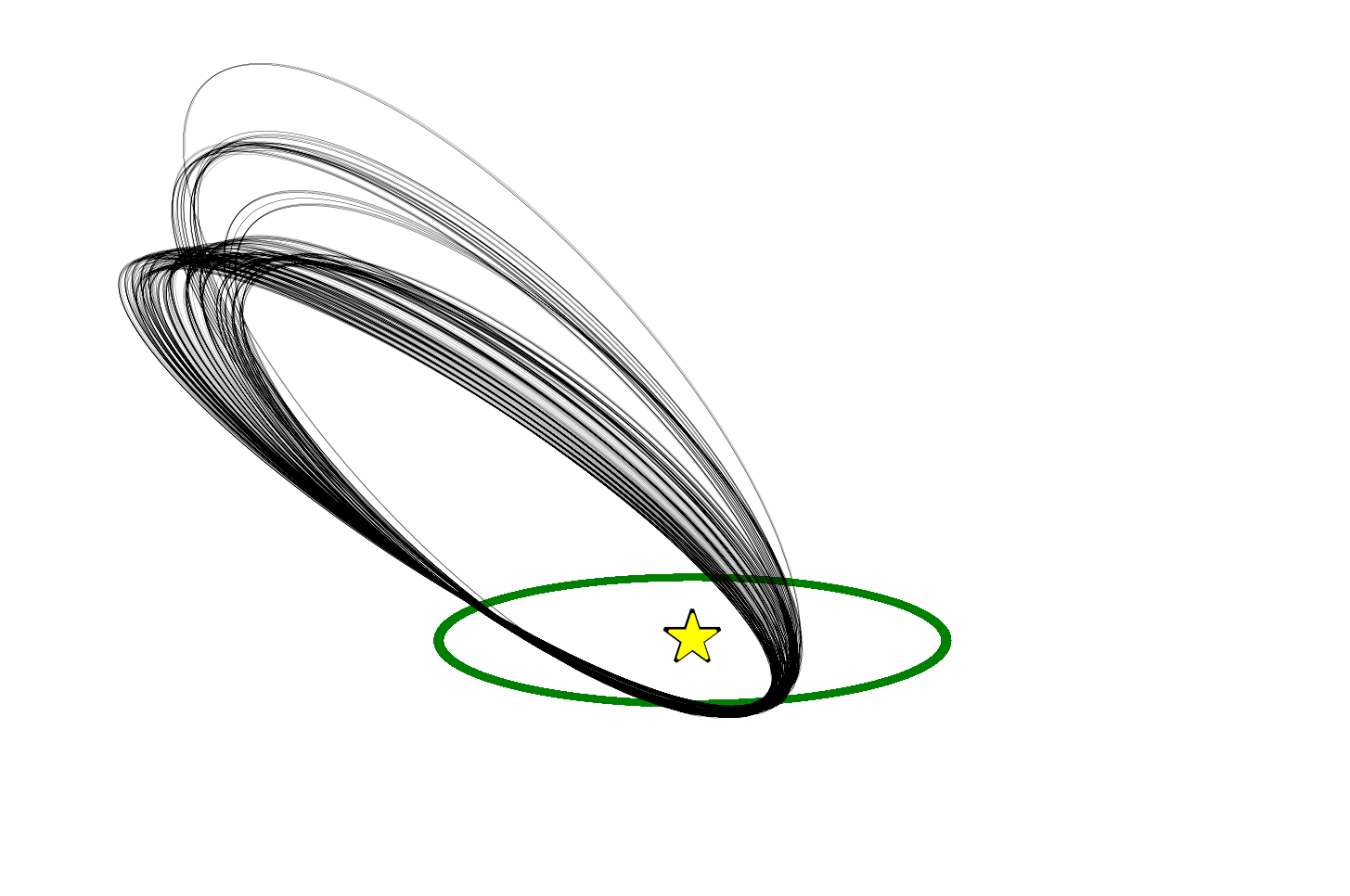}
    \caption{Left: numerical simulation of a single three-body capture of a test particle by the sun--Jupiter system. The simulation begins at the red dot. Right: long-term evolution of the captured object, showing successive changes in the orbital parameters.}
    \label{fig:capture-3d}
\end{figure}

In many cases of interest, the interaction of a binary with its environment can be reduced to a gravitational three-body problem. The phenomenology of the three-body problem is famously complex. In the case of scattering between a binary and a third body, the possible outcomes are similar to those of scattering between an atom and an external particle. In the simplest scenario, the third body can exchange energy with the binary, corresponding to excitation or relaxation of the atom in the microscopic analogy. But as with the analogy, there are more dramatic possibilities as well: the binary can be destroyed by the interaction, and one of the resulting components may form a new binary with the interloper. Alternatively, if the third body loses a significant amount of its kinetic energy without disrupting the binary, a new triple system may be formed, with the third body bound to the other two. This process is illustrated in \cref{fig:capture-3d}. Such triple systems are generally unstable. Eventually, the same three-body dynamics that allowed the third body to lose energy to the binary will allow one of the three bound bodies to gain energy from the other two and leave the system.

It is this sort of temporary capture which concerns us in this work. We are motivated by an apparently simple question: what are the properties of the population of captured objects in a given binary system? The resolution of this question is relevant to the study of free-floating exoplanets and their bound counterparts \citep{Smith:2001me,Hurley:2001wi,Perets:2012kk,2015MNRAS.449.3543W,2017ApJ...841...86B}, for example, but is also significant for less familiar objects. In particular, it is important for assessing the population of captured dark matter particles, or for characterising the demographics of compact objects that might be temporarily captured in observable binary systems, including the capture of interstellar objects in the solar system \citep{1983Obs...103....1V,1986AJ.....92..171T,1989MNRAS.241..507S,2003AsBio...3..207M,2018MNRAS.473.1589G,2018AJ....156..193L,2019MNRAS.487.3324G}.

The capture of unbound objects into bound orbits by binary systems has been studied by many authors in widely varying contexts. Three-body capture and ejection were studied systematically by \citet{1975MNRAS.173..729H}, who obtained approximate forms for the rates of these processes in cases where detailed balance can be applied. Subsequently, the theory of capture and ejection was extended by several authors to study comets in the solar system \citep{1983Obs...103....1V,1986AJ.....92..171T,1994Icar..108...18L,1999Icar..142..509D}, interstellar panspermia \citep{2003AsBio...3..207M,2018AJ....156..193L}, and the population of captured dark matter particles in the vicinity of Earth \citep{1987ApJ...321..571G,1988ApJ...328..919G,Lundberg:2004dn,Xu:2008ep,Peter:2009mm,Edsjo:2010bm}. A comprehensive account of results and astronomical applications is given by \citet{Valtonen:2005}.

Many of these studies are based on the results of detailed numerical simulations, which make it possible to study the properties of the captured population both immediately after capture and at late times. However, the results of such simulations are specific to the solar system. In scenarios involving extrasolar binary systems, it is important to have a simple description of capture processes that holds for a wide range of systems and interloper velocities. For such purposes, it is desirable to have a flexible semi-analytical framework for describing the population of captured objects---not only the capture and ejection rates, but also the distributions of orbital parameters of captured objects. Moreover, it is valuable to describe the dependence of each of these on the parameters of both the binary and the third body prior to capture. Finally, it is useful to obtain a simple geometric description of the types of encounters that lead to captures, and to understand the behavior of these captured trajectories at late times.

In this work, we develop such a formalism. We focus on captures resulting from a close encounter between a test particle and the smaller body of a binary, and we demonstrate that the form of the capture cross section in this case lends itself well to predictions of orbital parameters and ejection time-scales. In particular, within certain approximations, we show that the set of impact parameters leading to captures forms a disc whose parameters can be written in closed form. We use this result to derive analytical approximations for the capture rate, the orbital parameters of captured objects, and the ejection time-scale. Our results generalize those of \citet{1986AJ.....92..171T} and provide an analytical interpretation of the sorts of trajectories studied therein. We further extend the results to give a simple prescription for the ejection rate of captured objects as a function of their parameters upon capture.

This work is organized as follows. In \cref{sec:capture-cross-section}, we introduce our geometric formalism and estimate the capture cross section. In \cref{sec:capture-ejection-rate}, we use the same method to derive an analytical estimate of the ejection rate, and apply our results to estimate properties of the equilibrium population of captured dark matter particles. In \cref{sec:capture-numerical}, we compare our results with numerical experiments. We discuss our findings and conclude in \cref{sec:discussion}.

\section{The capture cross section}
\label{sec:capture-cross-section}

\begin{figure*}
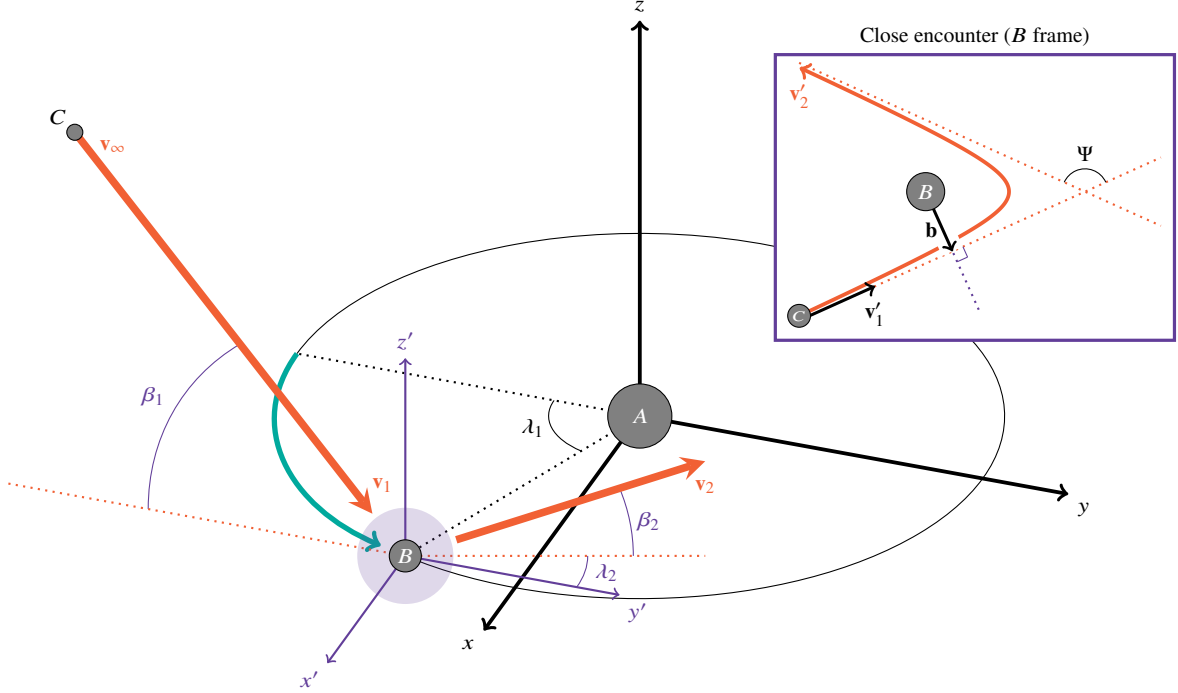
\centering
    % Colors
    \def\anglecolor{Emerald}
    \def\trajectorycolor{RedOrange}
    \def\primedcoordscolor{RoyalPurple}
    % Coordinates for 3d diagram
    \pgfmathsetmacro{\farawayi}{1.2}
    \pgfmathsetmacro{\farawayii}{0.7}
    \pgfmathsetmacro{\rvec}{.8}
    \pgfmathsetmacro{\betai}{50}
    \pgfmathsetmacro{\betaii}{20}
    \pgfmathsetmacro{\lambdai}{-20}
    \pgfmathsetmacro{\lambdaii}{110}
    \pgfmathsetmacro{\thetavec}{30}
    % Coordinates for 2d diagram
    \def\hypa{1}
    \def\hypb{0.45}
    \def\dommin{-2}
    \def\dommax{2}
    \def\arcrad{0.3}
    \def\focx{-\hypa-sqrt(\hypa^2+\hypb^2)}
    \def\focy{0}
    \def\asx{-\hypa*cosh(1)}
    \def\asy{-\hypb*sinh(1)}
    \def\pscale{0.15}
    \def\startx{-\hypa*cosh(\dommin)}
    \def\starty{\hypb*sinh(\dommin)}
    \def\blength{\hypb*(\hypa / sqrt(\hypa^2 + \hypb^2) + 1)}
    % Typeset the 2d diagram first
    \newsavebox{\twobodybox}
    \sbox{\twobodybox}{%
        \begin{tikzpicture}
            \input{two_body_encounter}
        \end{tikzpicture}
    }
    % Coordinate system for 3d diagram
    \tdplotsetmaincoords{60}{110}
    \begin{tikzpicture}[scale=6,tdplot_main_coords]
        \input{three_body_configuration}
        \node
            [fill=white, draw=\primedcoordscolor, very thick,
             label={Close encounter ($B$ frame)}]
            at (-0.5, 0.6, 0.4)
            {\usebox{\twobodybox}};
    \end{tikzpicture}
    \caption{Configuration and notation assumed in \cref{sec:capture-cross-section}. \textit{Centre:} 3d illustration of the encounter on the scale of the $AB$ system. Dotted lines lie in the plane of the $AB$ system. Notation largely follows \citet{1986AJ.....92..171T}. Object $C$, with velocity $\bb v_\infty$ at infinity, has a close encounter with object $B$ in the shaded region with initial velocity $\bb v_1$ and exits the encounter with velocity $\bb v_2$. Note that on the scale of the system as drawn, the trajectory of $C$ should be curved throughout due to acceleration by $A$, a feature we omit for simplicity. \textit{Inset:} 2d illustration of the close encounter in the frame of object $B$. Dotted lines lie in the plane of the two-body scattering process. The inset is intended only to illustrate the notation, and is not drawn to scale with respect to the centre image.}
    \label{fig:three-body-configuration}
\end{figure*}

Our goal is to identify the sorts of close encounters in which the incoming object is slowed enough to enter a bound orbit. In this section, we describe the set of impact parameters leading to captures, and connect this with both the capture cross section and the distributions of orbital parameters. We first establish our notation and approximations, which largely follow the presentation of \citet{1986AJ.....92..171T}. The notation is summarized in \cref{fig:three-body-configuration}.

We assume that the binary system is composed of two objects $A$ and $B$ with masses $M_A$ and $M_B$, and we take $M_A\gg M_B$. We use $\mu_X\equiv GM_X$ to denote the standard gravitational parameter for any object $X$, where $G$ is Newton's constant, and we denote the distance between any two objects $X$ and $Y$ by $r_{XY}$. While our formalism can be naturally extended to accommodate eccentric binaries, we take the orbit to be circular ($e=0$) in this work, so that $r_{AB}$ is constant. Unprimed quantities are measured in the frame of $A$ and primed quantities are measured in the frame of $B$. We assume that a test particle $C$ is incident from infinity with velocity $\bb v_\infty$, has a close encounter with object $B$, and thereafter becomes bound to object $A$. We write $\bb v_1$ and $\bb v_2$ to denote the velocity of $C$ just before and just after the close encounter.

The state of the binary is described by a single phase $\lambda_1$, and we define $\lambda_1=0$ to be the phase such that the $A$--$B$ axis is parallel to the projection of $\bb v_\infty$ in the plane of the orbit. We will assume that the time-scale of the close encounter is much smaller than the orbital time-scale of the $AB$ system so that $\lambda_1$ does not change significantly during the close encounter, i.e., we work in the impulse approximation. In general, $\bb v_\infty$ is inclined with respect to the orbital plane by an angle $\beta_1$, and $\bb v_2$ is inclined by an angle $\beta_2$. Additionally, we will speak of the impact parameter for the close encounter as a vector $\bb b$ in the frame of object $B$, spanning from $B$ to the point of closest approach of $C$ if the latter were to continue travelling undeflected with velocity $\bb v_1^\prime$ (see \cref{fig:three-body-configuration}, inset). We define $\bb b$ in the plane orthogonal to $\bb v_1^\prime$, endowing this plane with polar coordinates $(b,\phi)$. We will fix the axis $\phi=0$ shortly, and we will also return to the subtlety of frame-dependence in the definition of $\bb b$. First, however, we quantify the meaning of a close encounter.

For our purposes, a close encounter takes place when $C$ passes close enough to $B$ so that tidal acceleration by $A$ can be neglected. Then the encounter can be treated purely as a two-body problem in the frame of object $B$, greatly simplifying the analysis. This translates to the condition
\begin{equation}
    \frac{\mu_A}{(r_{AB} - r_{BC})^2} - \frac{\mu_A}{r_{AB}^2} <
        \epsilon\frac{\mu_B}{r_{BC}^2}
    \quad
    \text{for~some~}
    \epsilon\ll 1.
\end{equation}
Note that $r_{BC}$ is not a fixed parameter of the encounter, but rather evolves throughout the scattering process. The condition above determines which values of $r_{BC}$ are small enough to indicate a close encounter. To leading order in $\epsilon$, this condition can be written in the form
\begin{equation}
    r_{BC} \lesssim \rclose(\epsilon) \equiv
        r_{AB}\left(\frac{M_B\epsilon}{M_A}\right)^{1/3}.
\end{equation}
Note that $\rclose(\epsilon)$ is smaller than the Hill radius for $\epsilon\ll1$, and for a fixed choice of $\epsilon$, the value of $\rclose(\epsilon)$ defines what we mean by a close encounter. Later, when computing the capture cross section numerically, we will take $\epsilon=0.1$ and neglect trajectories for which $\min r_{BC}>\rclose(\epsilon)$. This leads to a conservative result for the capture cross section, but has the opposite effect on the ejection cross section, as we will discuss later. Since $M_A\gg M_B$, we will assume that $\rclose\ll r_{AB}$.

Having made this definition of a close encounter, we can compute $v_1$ as a function of $v_\infty$. Our approach assumes that the close encounter can be treated as an isolated two-body problem, which is only appropriate if the gravitational potential of object $B$ is small at $\rclose$. Otherwise, the acceleration of $C$ is dominated by the potential of $A$ for a significant part of the encounter, and by the time the two-body treatment is applicable, $C$ is already well within the potential of $B$. In the case that this effect can be neglected, it is sufficient to account for acceleration of $C$ by $A$ during infall from infinity to $\rclose$, which gives
\begin{equation}
    v_1 = \sqrt{v_\infty^2 + 2\mu_A/r_{AB}}.
\end{equation}
On the other hand, if the potential of $B$ is not small at $\rclose$, then $C$ has now been non-negligibly accelerated by $B$ prior to the close encounter, but $v_1$ must still be fixed where the close encounter begins. Thus, in general, we will include this additional prior acceleration, and we take
\begin{equation}
    \label{eq:v1-vinf}
    v_1 = \sqrt{v_\infty^2 + 2\mu_A/r_{AB} + 2\mu_B/\rclose(\epsilon)}
    .
\end{equation}
For the sun--Jupiter system, this additional acceleration contributes only a fraction of a percent to $v_1$, but in other realistic systems, the effect can be significantly larger. Note that this expression fixes only the speed $v_1$ in terms of $v_\infty$, and does not specify the vectorial relation between $\bb v_\infty$ and $\bb v_1$. We will return to the implications of directionality shortly.

Now, presuming a close encounter, we determine the conditions leading to capture of $C$. Under the stated assumptions, the relative velocity of $B$ and $C$ evolves as in the two-body problem from $\bb v_1^\prime$ to some $\bb v_2^\prime$. Object $C$ is bound after the close encounter if its speed is sufficiently low in the $A$ frame, i.e., if $v_2<\vesc$, where $\vesc=\sqrt{2\mu_A/r_{AB}}$ is the escape velocity of object $A$ at the location of the close encounter. The key feature of the two-body encounter for our purposes is that the speed of recession is equal to the speed of approach, i.e., $v_1^\prime = v_2^\prime$. This makes the outcome of the encounter very simple to describe analytically: the trajectory of $C$ is simply deflected by an angle $\Psi$ about the axis parallel to $\bb b\times\bb v_1^\prime$. The angle $\Psi$ is related to the impact parameter $b$ via
\begin{equation}
    \label{eq:deflection-angle}
    \cos\Psi = \frac{b^2v_1^{\prime4} - \mu_B^2}{b^2v_1^{\prime4} + \mu_B^2}
    .
\end{equation}
We can now compute $\bb v_2$ in terms of $\bb b$ algebraically. To be concrete, we first rotate the coordinate system so that $\bb v_1^\prime\propto\bb{\hat z}$ and $\bb b\propto\bb{\hat x}$ via a rotation $R_1$. Then the deflection of $\bb v_1^\prime$ into $\bb v_2^\prime$ is computed by performing a rotation by $\Psi$ in the $xz$-plane. This procedure allows us to define $\phi$ unambiguously: the impact parameter $\bb b$ lies in the plane orthogonal to $\bb v_1^\prime$, so in the rotated coordinate system, it takes the form $\bb b = (b_1,b_2,0)$. We define $\phi=0$ such that $\bb b$ lies in the $xy$-plane in the \emph{original} coordinate system. That is, we require that $(R_1^{-1}\bb b)\cdot\bb{\hat z}=0$. If we further choose that the $x$-component is positive, we can solve for $b_1$ and $b_2$ uniquely:
\begin{equation}
    \label{eq:base-impact-vector}
    \begin{pmatrix}
        b_1\\b_2
    \end{pmatrix}_{\phi=0} =
    \frac{b\sign\bigl(v_{1y}^\prime\bigr)}
        {\sqrt{v_{1x}^{\prime2} + v_{1y}^{\prime^2}}}
    \begin{pmatrix}
        v_{1y}^\prime \\ -v_{1x}^\prime
    \end{pmatrix}
    .
\end{equation}
Now $\bb b$ can be obtained for arbitrary $\phi$ by rotation of \cref{eq:base-impact-vector}.

To implement the deflection by $\Psi$, observe that in the new coordinate system, the rotation axis $\bb{\hat r}$ is related to $\bb b$ by a $\pi/2$ rotation. It is convenient to change coordinates with a rotation $R_2$ in the $xy$-plane to align $\bb b$ with the $x$-axis and $\bb{\hat r}$ with the $y$-axis. In the coordinate system produced by the rotation $R_2R_1$, the deflection corresponds to a simple rotation by $\Psi$ in the $xz$-plane, which we denote by $R_\Psi$. It follows that the deflection $\mathcal R\colon\bb v_1^\prime \mapsto \bb v_2^\prime$ is implemented by the matrix $\mathcal R = R_1^{-1}R_2^{-1}R_\Psi R_2R_1$. Using \cref{eq:deflection-angle} to write $\cos\Psi$ in terms of $b$, and using $\bb v_2=\bb v_2^\prime+\bb v_B$, we can now write $\bb v_2^\prime$ in terms of $b$. For brevity, we define
\begin{equation}
    a_{xy}\equiv\sqrt{a_x^2+a_y^2},
    \qquad
    q \equiv \sqrt{1 + \left.v_{1z}^{\prime2}\middle/v_{1xy}^{\prime2}\right.},
\end{equation}
for any vector $\bb a$. Then $\bb v_2$ is given by 
\begin{multline}
    \bb v_2 = \bb v_B +
        \left(\frac{b^2v_1^{\prime4} - \mu_B^2}
            {b^2v_1^{\prime4} + \mu_B^2}\right)\bb v_1^\prime\,+\\
        \frac{2\sign(v_{1y}^\prime)\mu_Bv_1^\prime b}
            {b^2v_1^{\prime4} + \mu_B^2}
        \begin{pmatrix}
            q\left(v_{1x}^\prime v_{1z}^\prime\sin\phi
                - v_1^\prime v_{1y}^\prime\cos\phi\right) \\
            q\left(v_{1y}^\prime v_{1z}^\prime\sin\phi
                + v_1^\prime v_{1x}^\prime\cos\phi\right) \\
            -v_1^\prime v_{1xy}^\prime\sin\phi
        \end{pmatrix}.
\end{multline}

Neglecting collisions with $B$, the condition for capture of $C$ can now be expressed succinctly as $v_2<\vesc$. Conveniently, it can be shown with some algebraic effort that saturation of this condition produces the equation of a circle in the plane of impact parameters. To this end, observe that the boundary relation $v_2-\vesc=0$ can be factored in the form
\begin{equation}
    v_2-\vesc =
        -F(\bb v_1,\bb b)\sign(v_{1y}^\prime)v_{1xy}^\prime v_1^{\prime4}
        \left(\frac{r_{AB}v_1^2 - 2\mu_B}{b^2v_1^{\prime4} + \mu_B^2}\right)
    ,
\end{equation}
for some factor $F(\bb v_1,\bb b)$. The remainder of the right-hand side depends on $\bb b$ only through the factor $b^2v_1^{\prime4}+\mu_B^2$, which is positive-definite. Thus, the right-hand side apart from $F(\bb v_1,\bb b)$ is non-zero almost everywhere, so our original condition can be rewritten in the form $F(\bb v_1,\bb b)=0$. Carrying out the factorization explicitly, $F$ has the form
\begin{equation}
    \label{eq:f-factor}
    F(\bb v_1,\bb b) = b^2 +
        \left(\frac{g_5}{g_4} - \frac{g_2}{g_1}\right) +
        b\left(\frac{g_3}{g_4}\cos\phi + \frac{g_1}{g_4}\sin\phi\right),
\end{equation}
where the coefficients $g_i$ are given in \cref{tab:coefficients}. In fact, the relation $F(\bb v_1,\bb b)=0$ is simply the equation of a circle in the plane orthogonal to $\bb v_1^\prime$, with radius $R$ and centre $\bb b_c$ given by
\begin{equation}
    \label{eq:capture-circle-parameters}
    R(\bb v_1) = \sqrt{
        \frac{g_2}{g_1} + \frac{g_1^2+g_3^2}{4g_4^2} - \frac{g_5}{g_4}
    }
    ,\qquad
    \bb b_c(\bb v_1) = -\frac{1}{2g_4}\begin{pmatrix}
        g_3\\g_1
    \end{pmatrix}
    .
\end{equation}
This allows us to make an extremely simple estimate of the capture cross section: we have simply
\begin{equation}
    \label{eq:sigma-capture}
    \scap(\bb v_1) \simeq
        \pi\min\left[R(\bb v_1),\;\rclose(\epsilon)\right]^2
        .
\end{equation}
When $R(\bb v_1)<\rclose(\epsilon)$, this takes the form
\begin{equation}
    \label{eq:sigma-capture-explicit}
    \scap(\bb v_1) \simeq \frac{\pi\mu_B^2\left[
        \left(v_1^{\prime2}-v_B^2\right)^2 - \vesc^4
    \right]}{
        \left(v_1^2-\vesc^2\right)^2v_1^{\prime4}
    }
    .
\end{equation}
This simple expression gives the capture cross section as a function of the incoming object's direction with respect to the axis of the binary---again, assuming a circular binary and working within the impulse approximation. When computing rates, the cross section should be multiplied by a factor of $v_1/v_\infty$ to account for gravitational focusing. Since $v_1$ and $v_1^\prime$ scale with $v_\infty$, the cross section vanishes rapidly for $v_\infty\gg\vesc$. On the other hand, as $v_\infty\to 0$, the velocity $v_1$ is nearly equal to $\vesc$, up to the small correction due to the potential of object $B$ (see \cref{eq:v1-vinf}). Thus, the cross section becomes very large, and is eventually subject to the cutoff in \cref{eq:sigma-capture}.

\begin{table}
    \hrule width \hsize \kern 1mm \hrule width \hsize height 1pt
    \begin{equation*}
        \renewcommand\arraystretch{2}
        \renewcommand\arraycolsep{1pt}
        \begin{array}{cl}
            g_1 &= 4\mu_B(\bb v_B\cdot\bb v_1^\prime) v_1^{\prime6}
                v_{1z}^\prime \\
            g_2 &= 4\mu_B^3(\bb v_B\cdot\bb v_1^\prime)v_1^{\prime2}
                v_{1z}^\prime \\
            g_3 &= 4\mu_Bv_1^{\prime7}
                \left(\bb v_1^\prime\times\bb v_B\right)_z \\
            g_4 &= \sign\bigl(v_{1y}^\prime\bigr)v_1^{\prime8}v_{1xy}^\prime
                    \left(v_1^2 - \vesc^2\right) \\
            g_5 &= 2\sign\bigl(v_{1y}^\prime\bigr)\mu_B^2v_1^{\prime4}
                v_{1xy}^\prime\left(v_1^{\prime2} + v_B^2 - \vesc^2\right)
            \\
        \end{array}
    \end{equation*}
    \hrule width \hsize \kern 1mm \hrule width \hsize height 1pt
    \caption{Coefficients $g_i$ appearing in \cref{eq:f-factor,eq:capture-circle-parameters}. Here $\bb v_1=\bb v_1^\prime+\bb v_B$ and $v_{1xy}^\prime$ is the magnitude of the projection of $\bb v_1^\prime$ onto the $xy$ plane.}
    \label{tab:coefficients}
\end{table}

\Cref{eq:sigma-capture-explicit} only holds for parameters such that $R(\bb v_1)$ is real in \cref{eq:capture-circle-parameters}, which is a non-trivial constraint. In particular, there is a maximum change in velocity that can be imparted to object $C$ during the encounter: the speed of approach is equal to the speed of recession in the frame of object $B$, so the maximum impulse corresponds to the case in which the direction of object $C$ is exactly reversed in the frame of $B$ (i.e., $\cos\Psi = -1$). In this case, $|\Delta v| = 2v_B$ in the frame of $A$. This means that there is a maximum velocity $v_{\mathrm{max}} = v_{\mathrm{esc}} + 2v_B$ such that objects with $v_1 > v_{\mathrm{max}}$ cannot be captured regardless of impact parameter. Such velocities correspond to non-real values of $R(\bb v_1)$, and for these velocities, the capture cross section is exactly zero.

We may now average over the binary phase $\lambda_1$ and arrival angle $\beta_1$ to obtain the directionally averaged cross section $\overline\scap$. Note that we use an overbar to indicate the directional average, reserving $\langle\cdot\rangle$ for the average over speeds. This requires care, however: not all arrival directions are kinematically allowed for fixed $v_1$ and $\vesc$, and it is difficult to analytically integrate only over parameters for which the expression of \cref{eq:sigma-capture-explicit} is positive-definite. Explicitly, the directional average should be computed by an integral of the form
\begin{equation}
    \overline\scap(\bb v_1) = \int\frac{\du\lambda_1}{2\pi}\dd\cos\beta_1\,
        \scap(\bb v_1)\,\chi(\bb v_1,\lambda_1,\beta_1)
    ,
\end{equation}
where $\chi$ is an indicator function equal to one when the arguments are kinematically allowed and zero otherwise. This average is readily carried out numerically, but $\chi$ is difficult to represent in closed form. However, for simplistic estimates, we can obtain an order-of-magnitude calculation of $\overline\scap$ by integrating over all arrival directions, including non-physical directions. We denote this quantity by $\widetilde\scap$, and it takes the form
\begin{multline}
    \label{eq:sigma-simple-average}
    \widetilde\scap\equiv
        \pi\left(\frac{\mu_B}{v_1^2-\vesc^2}\right)^2\Biggl[
            -1 - \left(
                \frac{\vesc^2 - v_B^2}{v_1^2 - v_B^2}
            \right)^2 +
            \\
            \frac{\vesc^2 + v_B^2}{v_1v_B}\arctanh\left(
                \frac{2v_1v_B}{v_1^2 + v_B^2}
            \right)
        \Biggr]
    .
\end{multline}
This is by no means a precise calculation, but the result is nonetheless quite useful, particularly for exhibiting the parametric dependence of the capture cross section on the binary configuration. The approximation breaks down most severely when $v_\infty$ is so small that $v_1\sim v_B$, but it is quite effective for larger values of $v_\infty$. For the sun--Jupiter system, we find $\widetilde\scap = 7.9A_J$ for $v_\infty=\SI{20}{\kilo\meter\per\second}$, where $A_J$ is the cross-sectional area of Jupiter. Full numerical integration over kinematically allowed angles gives $\overline{\scap}=9.9A_J$. To illustrate the applicability of this approximation, we compare the approximate and numerical results for several configurations in \cref{tab:approx-comparison}.

\begin{table}\centering
    {\renewcommand{\arraystretch}{1}\setlength{\tabcolsep}{8pt}
    \hrule width \hsize \kern 1mm \hrule width \hsize height 1pt
    \vspace{0.1cm}
    \begin{tabular}{ccccc}
        $r_{AB}$ & $M_B$ &
            $v_\infty\;[\SI{}{\kilo\meter\per\second}]$ &
            $\widetilde\scap\;[A_J]$ &
            $\overline\scap(\bb v_1)\;[A_J]$
        \\[0.1cm]\hline
        \\[-0.2cm]
        $r_{SE}$ & $M_E$ & 46.28 & \num{2.78e-6} & \num{3.50e-6} \\
        $r_{SJ}$ & $M_J$ & 20.23 & 7.133 & 9.074 \\
        $r_{SN}$ & $M_N$ & 8.436 & 0.732 & 0.924 \\[0.2cm]
        $r_{SE}$ & $M_J$ & 46.15 & 0.263 & 0.335 \\
        $r_{SJ}$ & $M_N$ & 20.28 & \num{2.19e-2} & \num{2.77e-2} \\
        $r_{SN}$ & $M_E$ & 8.439 & \num{2.51e-3} & \num{3.16e-3} \\[0.2cm]
        $r_{SE}$ & $M_N$ & 46.26 & \num{8.10e-4} & \num{1.02e-3} \\
        $r_{SJ}$ & $M_E$ & 20.29 & \num{7.52e-5} & \num{9.47e-5} \\
        $r_{SN}$ & $M_J$ & 8.417 & 238 & 303
        \\[0.2cm]
    \end{tabular}
    \hrule width \hsize \kern 1mm \hrule width \hsize height 1pt
    }
    \caption{Approximate and numerically averaged cross-sections for several configurations of object $B$. In each case, the velocity $v_\infty$ of the incoming object is fixed such that $v_1 = \frac12 v_{\mathrm{max}}$, where $v_{\mathrm{max}}$ is the maximum velocity with non-zero capture cross section (see text). This is chosen only as a representative velocity for typical captures. The subscripts $E$, $J$, and $N$ refer to Earth, Jupiter, and Neptune, respectively. For $X\in\{E,J,N\}$, $M_X$ denotes the mass of the planet $X$, and $r_{SX}$ denotes the distance between the Sun and the planet. The approximate cross section $\widetilde\scap$ slightly underestimates $\overline\scap$ by a consistent factor across configurations with widely varying parameters.}
    \label{tab:approx-comparison}
\end{table}

We can further directly obtain the differential cross section for a fixed specific energy transfer $\Delta\mathcal E\equiv\Delta E_C/M_C$. Since the potential energy is nearly the same immediately before and after the close encounter, we have $\Delta\mathcal E \approx \frac12(v_2^2-v_1^2)$, and thus we need only substitute $v_2(\mathcal E_2)$ for $\vesc$ in \cref{eq:sigma-capture-explicit}. This gives the total cross section to final states with specific energy below $\mathcal E_2$. Differentiating the resulting expression with respect to $\mathcal E_2$, and writing $\mathcal U(\epsilon) = -\mu_A/r_{AB}-\mu_B/\rclose(\epsilon)$, we find
\begin{multline}
    \label{eq:dsigma-de2}
    \frac{\du\scap(\bb v_1)}{\du\mathcal E_2} = \frac{
            16\pi\mu_B^2\left[\mathcal E_2+\mathcal U(\epsilon)\right]
        }{
            v_1^{\prime4}\left(
                v_1^2 - 2\left[\mathcal E_2 + \mathcal U(\epsilon)\right]
            \right)^3
        }\,\times\\
        \left[
                v_B^2\left(2v_1^{\prime2} + \bb v_B\cdot\bb v_1^\prime
            \right)
            +
            \left(
                v_1^{\prime2} -
                    2\left[\mathcal E_2 + \mathcal U(\epsilon)\right]
            \right)\left(
                \bb v_B\cdot\bb v_1^\prime
            \right)
        \right]
    ,
\end{multline}
as long as $R(\bb v_1,\mathcal E_2) < \rclose(\epsilon)$. Otherwise, while the desired specific energy transfer may not be kinematically prohibited, it cannot be attained by a two-body encounter with the specified value of $\epsilon$. We can approximate the directional average of this expression by starting instead with \cref{eq:sigma-simple-average}, which yields
\begin{multline}
    \label{eq:dsigma-de2-averaged}
    \frac{\du\widetilde\scap(v_1)}{\du\mathcal E_2} = 
    \frac{\pi\mu_B^2}{(\mathcal E_2-\mathcal E_1)^3}\Biggl[
        1
        + \frac{\mathcal E_2 - \mathcal E_1}
            {2[\mathcal E_1-\mathcal U(\epsilon)] - v_B^2} \\
        - \frac{\mathcal E_1 + \mathcal E_2 - 2\mathcal U(\epsilon) + v_B^2}
            {2v_B\sqrt{2[\mathcal E_1-\mathcal U(\epsilon)]}}
        \arctanh\left(
            \frac{2v_B\sqrt{2[\mathcal E_1-\mathcal U(\epsilon)]}}
                {2[\mathcal E_1-\mathcal U(\epsilon)] + v_B^2}
        \right)
    \Biggr]
    .
\end{multline}

Our computations thus far neglect the possibility of collisions with object $B$. In principle, it is possible that collisions also contribute to captures for compact objects such as light black holes. However, the relevant physics is quite different: energy is lost dissipatively by deformation of object $B$. For most cases of interest, the capture cross section is much larger than the collision cross section, but it is a simple matter to compute and subtract the latter if desired. The eccentricity $e_1^\prime$ and semimajor axis $a_1^\prime$ of the two-body hyperbolic orbit in the frame of object $B$ are given by
\begin{equation}
    e_1^\prime = \sqrt{1+ \frac{b^2v_1^{\prime4}}{\mu_A^2}}
    ,\qquad
    a_1^\prime = -\frac{b}{\sqrt{e_1^{\prime2} - 1}}
    .
\end{equation}
Then the pericentre is given by $r_{\mathrm{min}}=a_1^\prime(1-e_1^\prime)$, or
\begin{equation}
    r_{\mathrm{min}} = 
    \frac{\sqrt{\mu_A^2 + b^2v_1^{\prime4}} - \mu_A}{v_1^{\prime2}}
    .
\end{equation}
Requiring $r_{\mathrm{min}}>r_B$, we obtain the condition
\begin{equation}
    \label{eq:collision-radius}
    b > b_{\mathrm{min}} \equiv
    \frac{1}{v_1^\prime}\sqrt{2\mu_Br_B+\left(r_Bv_1^{\prime}\right)^2}
    .
\end{equation}
The set of impact parameters leading to collisions is, of course, also a circle. We can now write the cross section for captures without including collisions by simply subtracting the area of intersection of the two circles from our prior result. This is given by 
\begin{multline}
    \sigma_{\mathrm{int}} = -\biggl[
        \frac12
        \left(-b_c+R+b_{\mathrm{min}}\right)\times \\
        \left(b_c+R-b_{\mathrm{min}}\right)
        \left(b_c-R+b_{\mathrm{min}}\right)
        \left(b_c+R+b_{\mathrm{min}}\right)
    \biggr]^{1/2} + \\
    R^2\arccos\left(\frac{b_c^2 + R^2 - b_{\mathrm{min}}^2}{2b_cR}\right) + 
    r^2\arccos\left(
        \frac{b_c^2 + b_{\mathrm{min}}^2-R^2}{2b_cb_{\mathrm{min}}}\right)
    .
\end{multline}
In general, $\sigma_{\mathrm{int}}$ can be subtracted from $\scap$ to exclude collisions from the cross section. For our present purposes, we neglect the possibility of collisions altogether, so we do not carry out this subtraction in our subsequent results.

We can now use the capture cross section in \cref{eq:sigma-simple-average} to estimate the capture rate of test particles with velocity $v_\infty$ far from the binary system. First, however, it is necessary to convert $\scap(v_1)$ to the cross section $\scap(\bb v_\infty)$ pertinent to the rate calculation. The relationship between $v_1\equiv\|\bb v_1\|$ and $v_\infty\equiv\|\bb v_\infty\|$ is specified by \cref{eq:v1-vinf}. But the arrival direction of object $C$ at object $B$ is also influenced by acceleration due to object $A$, so the relationship between $\bb v_1$ and $\bb v_\infty$ has a non-trivial angular dependence. However, we expect this effect to have only a small impact on the directionally averaged cross-section: any modifications to $\lambda_1$ must disappear from the time-averaged cross section by azimuthal symmetry, so the sole effect of such deflection is to change the distribution of inclination angles $\beta_1$ of incoming objects. We are already treating this distribution crudely by integrating over non-physical arrival angles in \cref{eq:sigma-simple-average}, so we neglect this additional deflection, assuming that $\bb v_1\propto\bb v_\infty$.

With this assumption, we can write $\scap(\bb v_\infty) = \scap(\bb v_1(\bb v_\infty))$. Now, given a distribution function $f(\bb v_\infty)$ for the velocity at infinity, the capture rate can be estimated as $n\left\langle\scap v\right\rangle$, where $n$ is the number density of objects and the velocity-averaged cross section is given by 
\begin{equation}
    \left\langle\scap v\right\rangle =
        \int\du^3\bb v_\infty\, f(\bb v_\infty)\,
            \scap(\bb v_\infty)v_1(v_\infty)
    .
\end{equation}
Note the appearance of $v_1$ in place of $v_\infty$, accounting for the gravitational focusing factor $v_1/v_\infty$.

This formalism also lends itself well to describing the orbital parameters of captured objects. Since we have obtained $\bb v_2$ explicitly as a function of the impact parameter, we can readily compute the specific orbital energy $\mathcal E$ and specific angular momentum $\mathcal L$ of the captured object as
\begin{equation}
    \mathcal E_2 = \frac12\bb v_2^2 + \mathcal U(\epsilon),
    \qquad
    \mathcal L_2 = \left\|\bb r_{AB}\times\bb v_2\right\|,
\end{equation}
whereupon the eccentricity $e$ and semimajor axis $a$ of the captured object's orbit take the form
\begin{equation}
    e = \sqrt{1 + \frac{2\mathcal E_2\mathcal L_2^2}{\mu_A^2}},
    \qquad
    a = -\frac{\mu_A}{2\mathcal E_2}
    .
\end{equation}
The resulting expressions are algebraically complicated but are nonetheless tractable, and in closed form. Obtaining the full distributions of orbital parameters is analytically challenging, but readily performed semi-analytically: uniformly sampled points in the $(b,\phi)$ plane can now be converted to orbital parameters. In particular, we can evaluate $\bar e$ and $\bar a$ by numerically integrating over initial configurations which produce captures, i.e., over the circle described by \cref{eq:capture-circle-parameters}.

For an analytical estimate, we can translate \cref{eq:dsigma-de2-averaged} to an approximate differential cross section with respect to $a$, using
\begin{equation}
    \label{eq:semi-major-axis-distribution}
    \frac{\du\widetilde\scap(\bb v_1)}{\du a} = 
        \frac{\mu_A}{2a^2}
        \frac{\du\widetilde\scap(\bb v_1)}{\du\mathcal E_2}
    ,
\end{equation}
and thus obtain a probability distribution for $a$ as a function of $\bb v_1$. The binary is assumed to be circular, with fixed separation $r_{AB}$, and the captured orbit must cross the trajectory of object $B$, so we impose a lower cutoff $a>r_{AB}$. The resulting distribution is sharply peaked at small $a$, but does not have a well-defined mean. For comparison with numerical results, it suffices to evaluate $\bar a$ considering only captured orbits with $a<a_{\mathrm{max}}$. We denote this approximate mean by $\tilde a$. For instance, for the sun--Jupiter system with $v_\infty=\SI{20}{\kilo\meter\per\second}$, taking $a_{\mathrm{max}}=\SI{120}{\au}$ gives $\tilde a = \SI{15.5}{\au}$. This result is comparable to that described in fig. 5 of \citet{1986AJ.....92..171T}, although note that the latter gives an approximate result computed only for a fixed value of $\beta_1$. Alternatively, one can compute the median value of $a$, which is analytically challenging but readily performed numerically. For the aforementioned Solar system configuration, we estimate the median semimajor axis of captured objects at $\SI{13.7}{\au}$. The distribution of \cref{eq:semi-major-axis-distribution} is also in excellent agreement with numerical experiments, as we shall see in \cref{sec:capture-numerical}.

Estimating the eccentricity after capture is substantially more complicated, since the specific angular momentum is independent of the specific energy after capture. There is no obvious geometric structure to the final angular momentum, in contrast to the circular regions we have identified for the final energy, and in general, the average over arrival angles must be performed numerically. However, we can exploit the semimajor axis distribution to make a simplistic estimate, as follows. Generally $\bar a > r_{AB}$, but the orbit of object $C$ after capture must cross the orbit of object $B$. Thus, given a value of $a$, there is a minimal eccentricity $e_{\mathrm{min}}(a)$ needed to ensure that the perihelion of $C$ lies within the orbit of $B$, i.e., $a(1-e)<r_{AB}$. Saturating this condition gives the lowest possible eccentricity for a capture with a given value of the semimajor axis. In general, highly eccentric captures are possible at the extremes of the parameter space. Thus, for a first estimate of the orbital parameter distribution, we assume that eccentricity is uniformly distributed on $(e_{\mathrm{min}}(a),\,1)$ for fixed $a$. That is, we take
\begin{equation}
    \label{eq:eccentricity-distribution}
    \frac{\du^2\widetilde\scap(\bb v_1)}{\du a\dd e} =
        \frac{\du\widetilde\scap(\bb v_1)}{\du a}
        \frac{\Theta(1-e)\,\Theta(e-e_{\mathrm{min}})}{1-e_{\mathrm{min}}}
    ,
\end{equation}
where $\Theta$ is the Heaviside function. While crude, this is in reasonably good agreement with eccentricities extracted from numerical experiments, as we shall demonstrate in \cref{sec:capture-numerical}. We define a typical eccentricity $\tilde e(a)$ as the mean of the corresponding uniform distribution at fixed $a$, i.e., $\tilde e(a) = \frac12(1+e_{\mathrm{min}})$.

\begin{figure}\centering
    \includegraphics[width=\columnwidth]{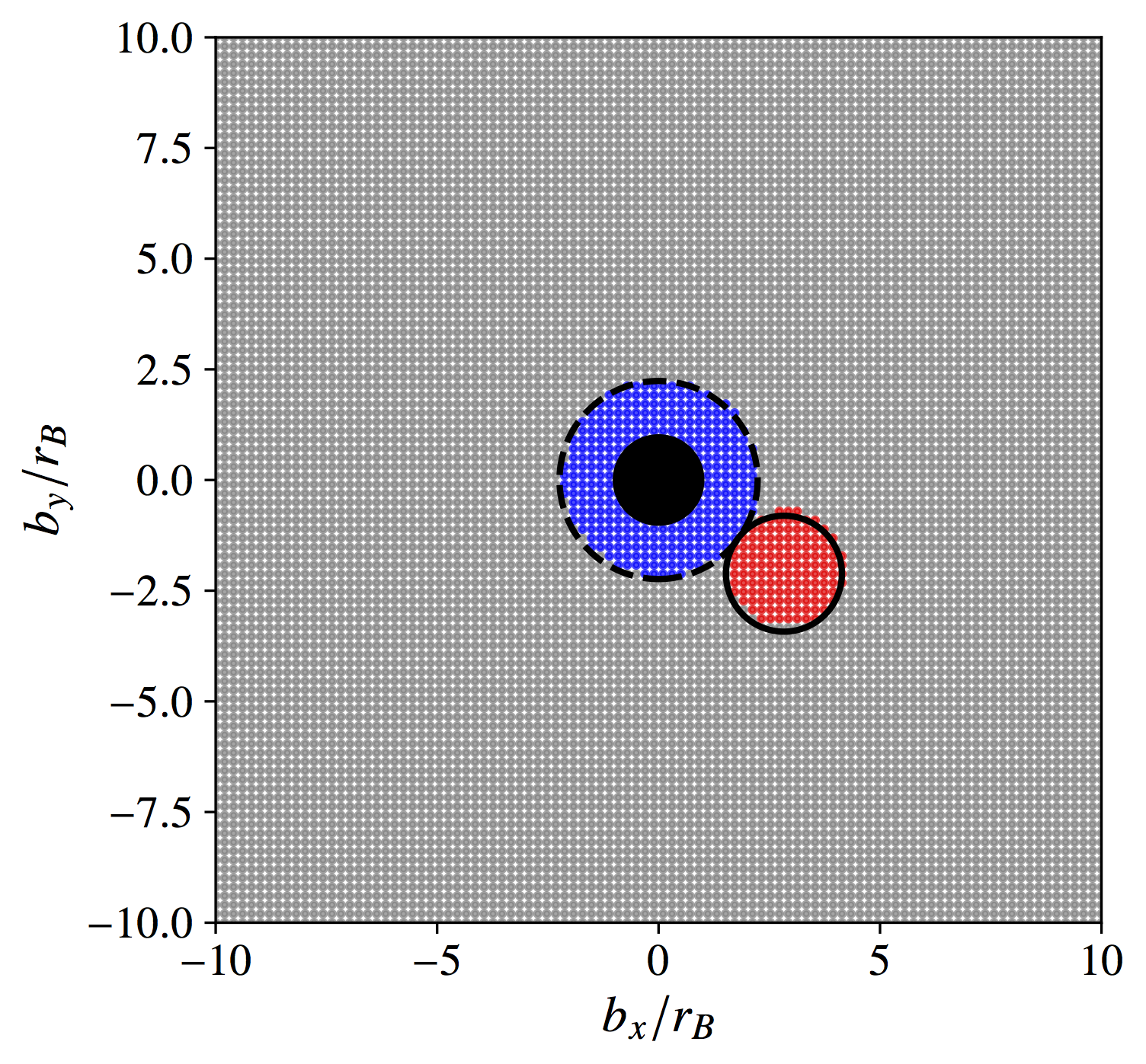}\\
    \includegraphics[width=\columnwidth]{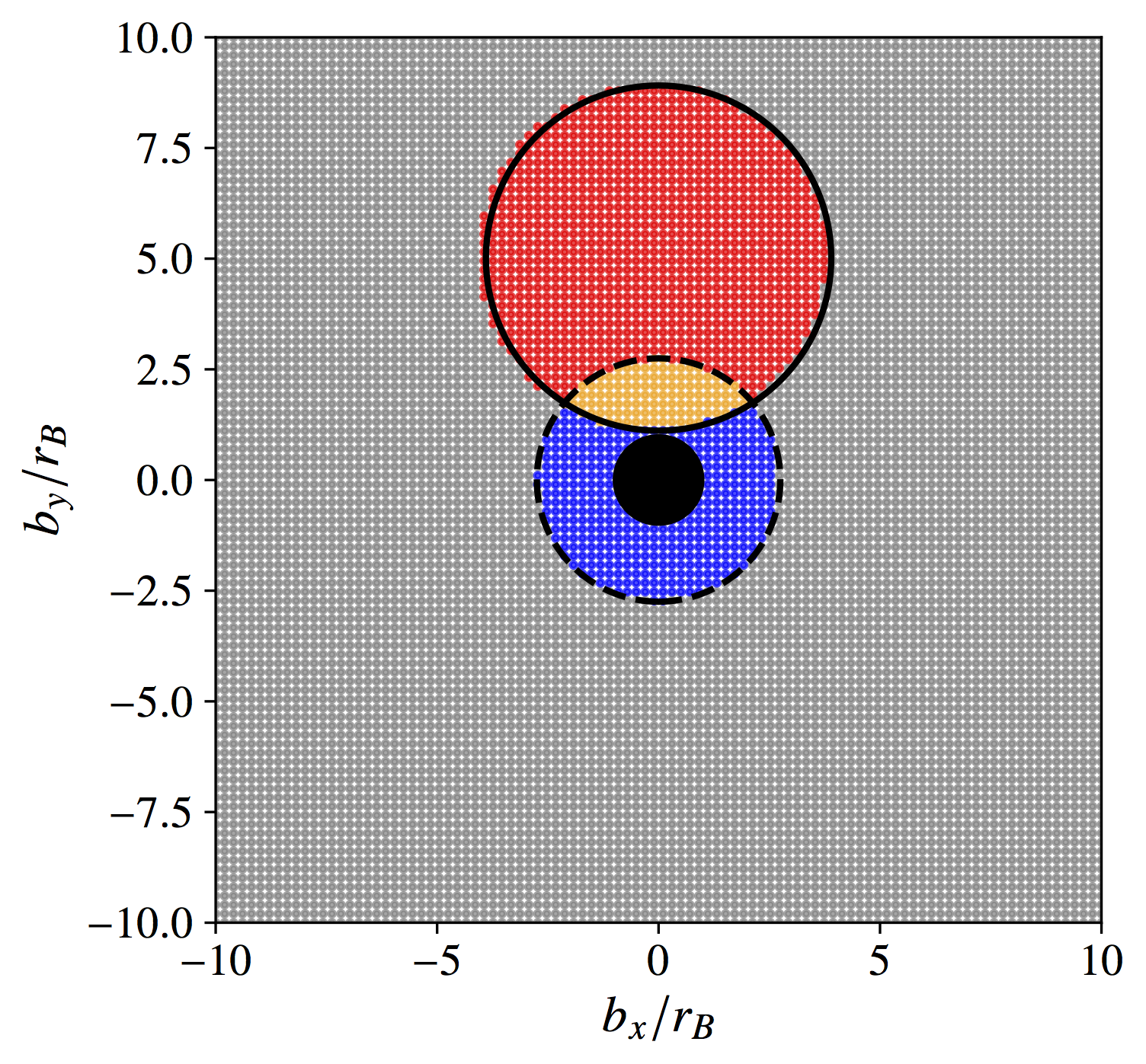}
    \caption{Outcomes of close encounters with Jupiter as determined by numerical integration, with $v_\infty=\SI{20}{\kilo\meter/\second}$, $\beta_1=\pi/3$, and two values of $\lambda_1$: in the top panel, $\lambda_1=0$, and in the bottom panel, $\lambda_1=\pi/2$. Each point represents an independent simulation with a different impact parameter. Points are shown in the plane of the impact parameter orthogonal to the velocity $\bb v_1^\prime$, i.e., from the perspective of object $C$ in the frame of object $B$. The angular coordinate is fixed by the prescription in \cref{sec:capture-cross-section}. Red points indicate capture of object $C$, and gray points indicate that object $C$ was unbound after departing from the close-encounter region. Orange points indicate that object $C$ would have been captured if Jupiter were replaced by a point mass, but instead suffered a collision. Blue points indicate collisions that would not yield captures even if Jupiter were compressed to a point. The solid circle shows the analytical prediction of the capture region in \cref{eq:capture-circle-parameters}, and the dashed circle shows the prediction of the collision region in \cref{eq:collision-radius}. Compare with fig. 4 of \citet{1986AJ.....92..171T}. Note that in the bottom panel, the red points are shifted very slightly to the left of the analytical prediction. This shift is in the direction of the sun and signals the presence of tidal forces.}
    \label{fig:impact-parameters}
\end{figure}

We now pause to compare our results to those of \citet{1986AJ.....92..171T} more generally. Figure~4 of that reference shows impact parameters leading to capture for several values of the orbital phase $\lambda_1$, similar to our \cref{fig:impact-parameters}. While the shape and position of each capture region is generally comparable to the circular region of \cref{eq:capture-circle-parameters}, there is clear distortion away from a circular shape. This is presumably due to one or both of two effects. One is our neglect of angular deflection between $\bb v_\infty$ and $\bb v_1$, but another is the definition of the impact parameter---and while the consequences for the capture rate are ultimately insignificant at the order-of-magnitude level, it is nonetheless important to understand the distinction between the two definitions.

Our formalism relies on the premise that the close encounter between objects $B$ and $C$ can be treated as a two-body encounter. Thus, working in the frame of object $B$, there is a natural definition of the impact parameter, which we temporarily denote by $\bb b^\prime$: it is simply the vector of closest approach between $B$ and the ray $\bb x_C^\prime\bigr|_{t=0} + \bb v_1^\prime t$ over all $t$. This is equivalent to the vector of closest approach between $B$ and $C$ in the absence of any interaction. The vector $\bb b^\prime$ is orthogonal to $\bb v_1^\prime$, but notice that it is \emph{not} orthogonal to $\bb v_1$, the initial velocity in the frame of object $A$. The impact parameter in the frame of $A$ has a different meaning. Indeed, in general, the magnitude of the impact parameter, as defined via the closest approach of the initial velocity ray to the second object, is only invariant between frames in which the initial velocities of $B$ and $C$ are parallel. The frame of $B$ is of course such a frame, but the frame of $A$ is generally not.

This means that any statements involving the impact parameter require us to specify the choice of frame. For our purposes, there are two relevant statements with such a dependence. One statement is the relationship of \cref{eq:deflection-angle} between the impact parameter and the deflection angle $\Psi$. This is formulated in the two-body problem, where the impact parameter is specified in a frame where the velocities are parallel. Thus, for calculation of the deflection angle, we must use the impact parameter $\bb b^\prime$, as calculated in the frame of $B$, and not its equivalent in frame $A$. The other statement concerns the relationship of the impact parameter to the cross section. Ultimately, the set of impact parameters that result in capture forms a region in the plane orthogonal to velocity whose area is the capture cross section. While the total cross section is the same between the frames of $A$ and $B$, the impact parameters are not, and thus, the shape of the capture region must transform in a complicated way to compensate.

We have checked that defining the impact parameter in the frame of $A$ produces regions in the impact parameter plane that more closely resemble the non-circular shapes of \citet{1986AJ.....92..171T}. In \cref{sec:capture-numerical}, we numerically validate our analytical prescription, and show that the capture regions are indeed circular under our stated assumptions and conventions.

\section{Estimating the ejection rate}
\label{sec:capture-ejection-rate}
In two-body dynamics, a pair of gravitationally bound objects remain bound forever. This is not the case in a three-body system for exactly the same reason that capture of the third body is possible: since the system is time-reversal invariant, the same process can take place in the opposite direction. A close approach between two bodies in a three-body bound system can transfer energy between them and lead to ejection of one of the two bodies from the system.

Unfortunately, estimating the rate of ejection from first principles is very challenging. As \citet{1975MNRAS.173..729H} explains, the complicated dynamics of the three-body system mean that the orbital configurations are constantly changing in an unpredictable fashion. The most reliable estimates of ejection time-scales come from direct numerical simulation of such systems, and even these are difficult to execute reliably over the potentially long time-scales involved. However, short of such a calculation, it is nonetheless useful to have an order-of-magnitude estimate of the lifetime of bound orbits under particular conditions. In the present context, our interest lies in estimating the statistics of the population of captured particles across a variety of systems \emph{without} expensive simulations, so it is useful to at least understand the basic dependence of the ejection rate on binary parameters.

In practice, ejection time-scales are often estimated using simplified Monte Carlo algorithms based on \"Opik theory \citep{1951PRIA...54..165O,1961AJ.....66..381O,1967JGR....72.2429W,1981Icar...48...39K} instead of full numerical simulations, an approach known as the \"Opik--Arnold method \citep{1965ApJ...141.1536A}. In our framework, since we can estimate the relevant cross-sections analytically, we can perform a semi-analytical analogue of the \"Opik theory estimate without any actual simulation. Since this approach is fundamentally rooted in the same approach as \"Opik--Arnold codes, we first review the typical algorithmic method.

 The \"Opik--Arnold estimate of the ejection rate relies on the assumption that the ejection process is driven by close encounters. The problem can then be decomposed into two parts: (1) determining the rate of close encounters, and (2) determining the outcome of each close encounter as it affects the orbital parameters of the captured object. \citet{1951PRIA...54..165O} estimates the time-scale between close encounters as a function of the orbital parameters of both objects, providing a solution to the first part of the problem. The second part can be approached iteratively via a Monte Carlo algorithm, randomly choosing an impact parameter for each close encounter and determining the new set of orbital parameters. While the algorithmic estimate is not in perfect agreement with numerical integration, it is capable of giving an inexpensive order-of-magnitude estimate of the ejection time-scale (see \citet{1999Icar..142..509D} for an extensive discussion).

However, despite the simplicity of the \"Opik--Arnold algorithm, it is inherently stochastic and iterative. This makes it difficult to produce straightforward analytical estimates of the ejection time-scale without a computational implementation. Thus, the primary advantage of the algorithmic approach is that it is much faster and simpler to implement than full numerical integration. For our purposes, however, we would like to have an order-of-magnitude estimate of the ejection rate that can be written in closed form, or at least evaluated semi-analytically. Our explicit algebraic results derived in the previous section make such a simplistic estimate possible, under the following assumptions:
\begin{enumerate}
    \item ejection of object $C$ is driven by close encounters, and
    \item close encounters take place mainly with object $B$.
\end{enumerate}
Note that since the initial orbital parameters of object $C$ are determined during a close encounter with object $B$, its initial orbit includes the point of the close encounter. It follows that the orbit of object $C$ crosses the orbit of object $B$, at least initially, justifying our second assumption.

There are now two strategies one could use to estimate the ejection time-scale. The first is to follow essentially the same strategy as the \"Opik--Arnold algorithm, but to use semi-analytical averages rather than iterative Monte Carlo computations. In particular, in the limit that there is a large number $N$ of close encounters prior to ejection, the specific energy transfer $\Delta\mathcal E$ can be treated differentially, writing $\du\mathcal E_C/\du N = \left.\left\langle\Delta \mathcal E\right\rangle\right|_{\mathcal E_C}$. In principle, using the differential cross section in \cref{eq:dsigma-de2-averaged}, one can explicitly evaluate $\left.\left\langle\Delta\mathcal E\right\rangle\right|_{\mathcal E_C}$, integrate this separable differential equation, and then solve $\mathcal E_C(N_{\mathrm{ej}})=0$ to determine the number $N_{\mathrm{ej}}$ of close encounters required to produce an ejection event. Once $\mathcal E_C(N)$ is obtained in closed form, one can approximate the time-scale between close encounters as a function of $\mathcal E_C$, and integrate on $N\in(0, N_{\mathrm{ej}})$ to finally estimate the ejection time-scale.

While certainly possible numerically, this process is algebraically formidable, and thus offers no great advantage over the \"Opik--Arnold treatment for an order-of-magnitude estimate. We therefore choose radical acceptance of our limitations, and propose an alternative method for an even simpler estimate of the ejection time-scale. While the orbital parameters of object $C$ certainly change significantly over the lifetime of the bound configuration, we make the following assumptions in addition to the previous two:
\begin{enumerate}
    \setcounter{enumi}{2}
    \item most close encounters do not substantially change the \emph{ejection cross section} in subsequent orbits, and
    \item most encounters at distance $r_1$ do not substantially change the time between subsequent close encounters at distances $r\ll r_1$.
\end{enumerate}
We caution that these assumptions are almost certainly flawed in most cases, but they may nonetheless suffice for a very simplistic parametric estimate. 

The value of these approximations, on the other hand, is significant: taken together, they imply that we may ignore all close encounters except those which lead directly to ejection. Given the ejection cross section, we can then use the same \"Opik formalism to estimate the rate of such close encounters, and thus produce an estimate of the ejection rate. In principle, neglecting distant encounters is not all that different from what is typically done in \"Opik--Arnold codes, which themselves neglect encounters falling beyond the influence radius of object $B$: implementations of the algorithm often include an enhancement factor alongside the cross section of the sphere of influence to account for the aggregate effects of such distant encounters. We do the same to a somewhat greater extent, as we will detail shortly.

Now all that remains is to compute the ejection cross section $\sej$. Fortunately, this much is easy in our formalism. The ejection cross section is simply the cross section for a close encounter with object $B$ in which the energy exchange is large enough that object $C$ becomes unbound, but apart from the amount of energy to be transferred, this is identical to the capture cross section, and we can thus use the same technology to compute the ejection cross section. In particular, \cref{eq:sigma-simple-average} holds in identical form, with $v_1$ replaced by $v_2$, the speed of object $C$ immediately after the close encounter leading to capture.

To implement this calculation, we follow the \"Opik-theory estimate of the close encounter time-scale as presented by \citet{1999Icar..142..509D}. With non-canonical units restored, the close encounter rate is given by 
\begin{equation}
    \label{eq:encounter-rate}
    \frac{\du N}{\du t} =
        \left(\frac{v_B\sqrt{r_{AB}}}{2\pi}\right)
        \frac{KW\tau^2}{\pi SW_xr_{AB}^2a^{3/2}}
    .
\end{equation}
Here $\tau$ is the length associated with the encounter cross section, i.e., $\sigma=\pi\tau^2$; $a$ is the semimajor axis of object $C$; $W$ is the approach speed, analogous to $v_1$ in the capture case; $W_x$ is the component of $\bb W$ parallel to $\bb r_{AB}$; and $K$ is the enhancement factor to the cross section mentioned previously, whose value we will address shortly. We determine $W$ and $W_x$ following \citet{1999Icar..142..509D},\ignorespaces
\footnote{Note that \citet{1999Icar..142..509D} denote our $W$ and $W_x$ by $U$ and $U_x$. We use $W$ to avoid confusion with $\mathcal U(\epsilon)$.}
and we likewise set $S = \max\left(\sin i,\;\tau / r_{AB}\right)$, where $i$ is the orbital inclination of object $C$.

We assume that the orbital parameters of object $C$ change rapidly enough on the time-scales of ejection that we may average over $i$. The average can be performed explicitly in terms of elliptic integrals, and since we may safely assume that $\sigma\ll r_{AB}^2$, the result simplifies to
\begin{multline}
    \label{eq:ejection-rate}
    \widetilde\Rej \simeq 
        \frac{Kv_B^2\widetilde\sej}{2\pi^{5/2}r_{AB}^{3/2}a^{3/2}\sqrt{W_x}}
        \times \\
        \Biggl\{
            2\sqrt\xi -
            \kappa_-\arctan\left(\frac{\sqrt\xi}{\kappa_-}\right) - 
            \kappa_+\arctan\left(\frac{\sqrt\xi}{\kappa_+}\right) + \\
                i\biggl[
                    \kappa_-\arctanh\left(\frac{\kappa_+}{\kappa_-}\right) +
                        \kappa_+\arctanh\left(1 +
                            \frac{\eta\widetilde\sej}
                                {2\pi r_{AB}^2\kappa_+^2}
                        \right)
                        -2i\kappa_+
                \biggr]
        \Biggr\}
    ,
\end{multline}
where for brevity we define
\begin{equation}
    \label{eq:ejection-shorthands}
    \eta=\sqrt{a(1-e^2)/r_{AB}}
    ,\quad
    \xi=3-r_{AB}/a
    ,\quad
    \kappa_\pm=\sqrt{-\xi\pm2\eta}
    .
\end{equation}
The ejection cross section can be written explicitly as 
\begin{multline}
    \label{eq:ejection-cross-section}
    \widetilde\sej = \pi\left(\frac{2M_Br_{AB}}{5M_A}\right)^2\Biggl[
        -1
        -\left(\frac{v_B^2r_{AB} - 2\mu_A}{2v_B^2r_{AB} + \mu_A}\right)^2 - \\
        \frac{v_B^2r_{AB} + 2\mu_A}{v_B\sqrt{\mu_Ar_{AB}/2}}\arctan\left(
            \frac{2v_B\sqrt{2\mu_Ar_{AB}}}{\mu_A - 2v_B^2r_{AB}}
        \right)
    \Biggr]
    .
\end{multline}
Taken together, \cref{eq:ejection-rate,eq:ejection-shorthands,eq:ejection-cross-section} allow for an analytical estimate of the ejection rate. We can certainly average the ejection rate over $a$ and $e$ values using the joint distribution of \cref{eq:eccentricity-distribution}. However, by simply substituting $\tilde a$ and $\tilde e(\tilde a)$ for $a$ and $e$, we obtain a crude but closed-form estimate for the typical lifetime of a captured orbit in a given binary system.

This estimate should be understood as an estimate of the mean of some distribution of lifetimes of captured orbits. The shape of this distribution reflects our assumption that close encounters can be treated as a Poisson process: if this were exactly true, the distribution of lifetimes $T$ would be exponential, with the probability distribution $f(T) = \Rej\exp\left(-\Rej T\right)$. This is potentially complicated by the effects of other close encounters: in principle, as in the \"Opik--Arnold approach, the trajectory of a typical capture is influenced by several other close encounters before the one which leads directly to ejection. If ejection is modelled as the cumulative outcome of some $N$ close encounters, each of which takes place with a comparable time-scale $T_1$, then the lifetime is distributed as a sum of $N$ exponentially distributed random variables, i.e., according to the Erlang distribution $E(N,T_1^{-1})$. Thus, the shape of the lifetime distribution is a key test of our simplistic ejection model: an exponential distribution is compatible with our assumptions, while a more general Erlang distribution signals the non-trivial involvement of multiple close encounters. In \cref{sec:capture-numerical}, we will see that the distribution of lifetimes in numerical experiments is well-fit by an exponential distribution, justifying the assumptions of this section.

With a complete estimate in hand, we can now compare to numerical benchmarks to estimate an appropriate value for $K$. We will carry this out in detail in \cref{sec:capture-numerical}, but for the moment, we note that $K\sim25$ is appropriate for order-of-magnitude estimates. As expected, this is somewhat larger than the value $K\sim 10$ preferred by \"Opik--Arnold codes to account for encounters lying beyond the influence radius.

Having developed a set of analytical approximations for the rates of capture and ejection, we now turn to the properties of the equilibrium population: in the limit of long times, what is the expected number of captured objects bound to object $A$? In equilibrium, the ejection rate balances the capture rate. Now, if the captured objects do not interact among themselves, then the capture rate is independent of the number of captured objects, while the ejection rate is proportional thereto. Thus,
\begin{equation}
    \overline N = \Rcap / \Rej
    .
\end{equation}
We can thus estimate $\overline N$ by $\widetilde N\equiv\widetilde\Rcap/\widetilde\Rej$ for fixed $v_\infty$. If the population of free objects interacting with the binary has a distribution $f(v_\infty)$, then we can average over the population and write
\begin{equation}
    \label{eq:equilibrium-population}
    \langle\widetilde N\rangle = n_\infty\int\du v_\infty\,f(v_\infty)
        \frac{\widetilde\scap(v_\infty)v_1(v_\infty)}{\widetilde\Rej(v_\infty)}
    ,
\end{equation}
where $n_\infty$ is the number density far from the binary. In general, this integral must be performed numerically. Nonetheless, this procedure allows for a rapid order-of-magnitude estimate of the equilibrium number of captured objects.

To demonstrate, we apply this method to the capture of particle dark matter with no non-gravitational interactions. This scenario has been studied extensively for the case of the solar system \citep{1987ApJ...321..571G,1988ApJ...328..919G,Lundberg:2004dn,Peter:2009mm}, so we likewise make an estimate for the sun--Jupiter system. We can make a simple semi-analytical estimate using an isotropic Boltzmann distribution for $f(v_\infty)$, i.e., neglecting the dark matter wind. Such a distribution has the form $f(v_\infty) \propto v_\infty^2\operatorname{e}^{-v_\infty^2/v_0^2}$, so that $f(v_\infty)\sim v_\infty^2/v_0^3$ at low velocities, with an exponential cutoff for $v_\infty\gtrsim v_0$. Note that $v_0$ for the local dark matter distribution is much larger than the orbital speed of Jupiter, so the low-velocity tail dominates the capture rate. We can numerically evaluate \cref{eq:equilibrium-population}, taking $\widetilde\scap$ from \cref{eq:sigma-simple-average}, $v_1(v_\infty)$ from \cref{eq:v1-vinf}, and $\widetilde\Rej$ from \cref{eq:ejection-rate}. Taking an rms velocity of \SI{220}{\kilo\meter\per\second} for the dark matter particles, we find $\langle\widetilde N\rangle \simeq (\SI{0.1}{\au^3})n_\infty$. Compared to the number density $n_\infty$ in the spherical volume within Jupiter's orbit, this corresponds to an $\mathcal O(10^{-4})$ enhancement. This is reasonably consistent with detailed simulations by \citet{Peter:2009mm}, who finds that the density enhancement at Earth is sub-per cent.

\section{Comparison with numerical integration}
\label{sec:capture-numerical}
In the previous section, we obtained analytical results for the capture cross section, and semi-analytical results for the distribution of orbital parameters. These results are only reliable within the context of the stated approximations, and it is thus important to compare them with numerical results to be assured of their validity in the regimes of interest. We will begin our numerical analyses with the sun--Jupiter system, since this system has been extensively studied by prior authors, and thus serves as a well-understood benchmark.

We numerically integrate the equations of motion using the \textsc{mercurius} integrator \citep{2019MNRAS.485.5490R} via the publicly-available \textsc{rebound} code \citep{2011ascl.soft10016R,2012A&A...537A.128R}. In each simulation, we configure the three bodies $A$, $B$, and $C$ according to fixed values of $\lambda_1$, $\beta_1$, and $v_1$. We set the initial position of object $C$ in the frame of object $B$, offset by a vector of length $\rclose(\epsilon=0.1)$ in the direction of $-\bb v_1^\prime$ and by an orthogonal vector $\bb b$. In the following, we shall describe $\bb b$ as a 2d vector in the plane orthogonal to $\bb v_1^\prime$. We always fix $v_\infty$ and derive $v_1^\prime$ from \cref{eq:v1-vinf} to avoid unphysical speeds.

We begin with the dynamics of captures. Our first goal is to confirm our statements regarding the \emph{shape} of the capture region in the plane of the impact parameter $\bb b$. To that end, we configure simulations with varying impact parameter $\bb b$, and for each such configuration, we test whether $C$ becomes bound to the sun before leaving the close-encounter region. We diagnose a capture trajectory as one for which object $C$ is initially free, i.e., $\mathcal E_1>0$, and for which object $C$ becomes bound to object $A$ at some later time, i.e., $\frac12v_C^2-\mu_A/r_{AC}<0$ in the frame of object $A$. \Cref{fig:impact-parameters} shows the results of our numerical simulations for the same parameters used in fig. 4 of \citet{1986AJ.....92..171T}, demonstrating excellent agreement with our analytical predictions. Note that the impact parameter used in \cref{fig:impact-parameters} is defined as in \cref{sec:capture-cross-section}. As a benchmark, the capture cross section for objects with $v_\infty=\SI{20}{\kilo\meter\per\second}$ and inclination $\beta_1=\pi/3$ is $4.8A_J$, where $A_J$ is the cross-sectional area of Jupiter. This agrees with the result of \citet{1986AJ.....92..171T}, who finds this cross section to be ``roughly five times the area of Jupiter.''

\begin{figure}\centering
    \includegraphics[width=\columnwidth]{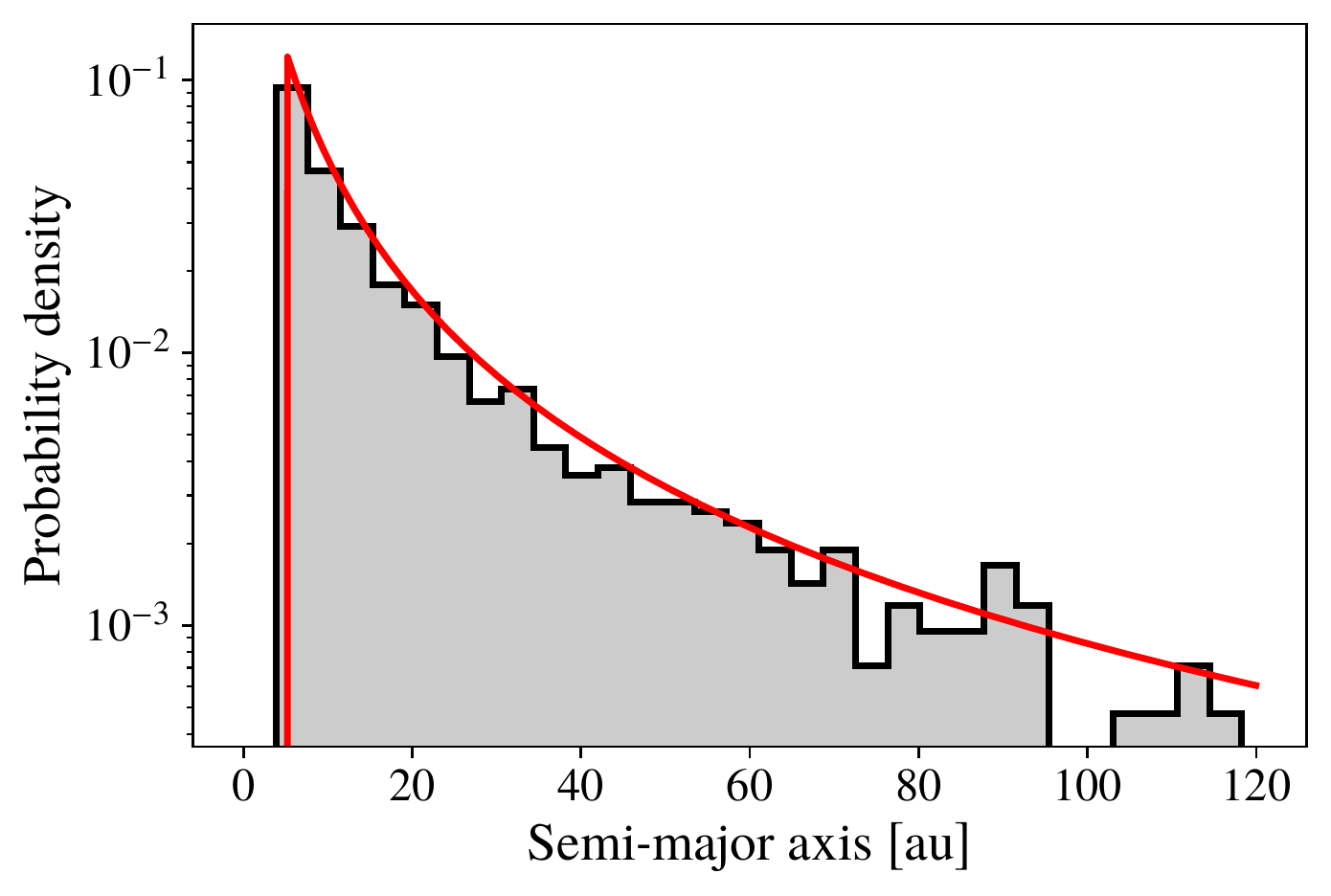}\\
    \includegraphics[width=\columnwidth]{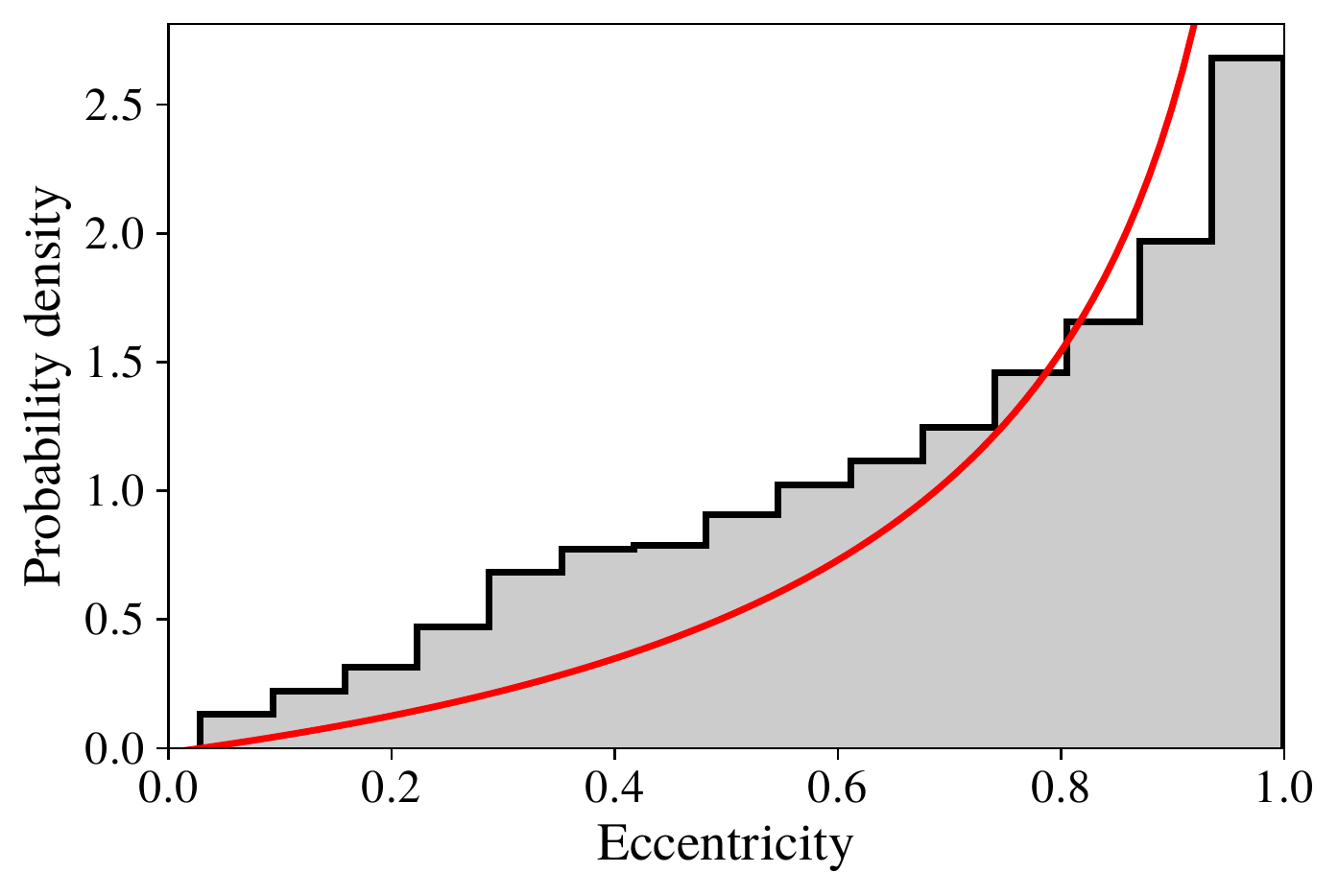}
    \caption{Top: distribution of semimajor axes immediately after capture by the sun--Jupiter system for $v_\infty=\SI{20}{\kilo\meter/\second}$. The histogram shows the distribution extracted from an ensemble of simulations (see text for details). The red line shows the prediction of \cref{eq:semi-major-axis-distribution}. Bottom: distribution of eccentricities. The solid red line shows the prediction of \cref{eq:eccentricity-distribution}, marginalizing over $a$.
    }
    \label{fig:orbital-parameters-prompt}
\end{figure}

We compare analytical predictions of the orbital parameter distributions to numerical results in \cref{fig:orbital-parameters-prompt}. The analytical semimajor axis distribution is in good agreement with numerical results. Our estimate of the eccentricity distribution is very crude, based only on heuristic arguments, but it nonetheless traces the essential behavior of the numerical results. We stress that these orbital parameters are not time-invariant, but evolve after the capture. This is a key difference between two-body and three-body dynamics. \Cref{fig:orbital-parameters-prompt} shows the orbital parameters only immediately after capture.

\begin{figure}
    \includegraphics[width=\columnwidth]{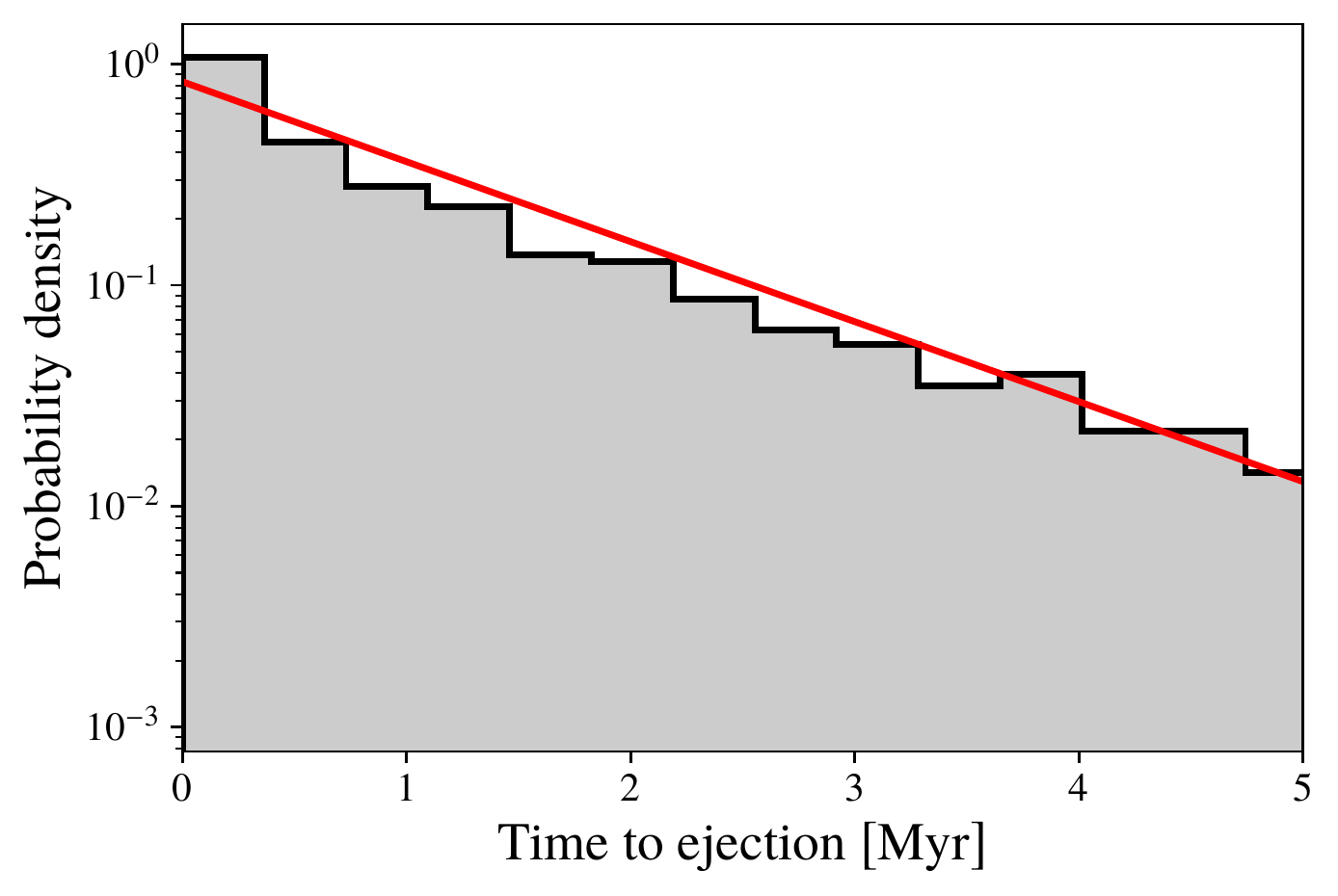}
    \caption{Distribution of capture lifetimes (i.e., time from capture to ejection) in an ensemble of 2500 simulations in the sun--Jupiter system with $v_\infty=\SI{20}{\kilo\meter/\second}$. The red curve shows an exponential distribution with the estimated ejection rate of \cref{eq:ejection-rate} ($K=25$).}
    \label{fig:lifetime-distribution}
\end{figure}

Finally, we test our prediction of the ejection time-scale against numerical integration. For the sun--Jupiter system with $v_\infty=\SI{20}{\kilo\meter/\second}$, our prescription estimates the typical ejection time-scale at $1/\widetilde\Rej=(\SI{3.0e7}{\year})/K$. We determine the mean ejection time-scale numerically by integrating an ensemble of initial conditions, randomly sampled with isotropic arrival directions and with impact parameters sampled uniformly in the plane orthogonal to $\bb v_1^\prime$. As in \cref{fig:impact-parameters}, we include impact parameters that lie outside the capture region according to our analytical prediction, but we discard all configurations which do not result in capture of object $C$. We integrate forward in time until object $C$ is ejected. This ensemble of simulations gives the mean ejection time-scale as $\bar t_{\mathrm{ej}} \simeq \SI{1.2e6}{\year}$, suggesting $K\sim 25$, as noted in \cref{sec:capture-ejection-rate}. A very small number of initial conditions lead to long-lived captures that are not ejected within the running time of our simulations, and the impact of these points in our subsequent analysis is negligible.

It is certainly encouraging that our analytical estimate can reproduce numerical results with a value of $K$ only an $\mathcal O(1)$ factor larger than that used in \"Opik--Arnold codes. A larger value of $K$ is expected, of course---our analytical estimate neglects contributions from a larger set of close encounters than are neglected in the \"Opik--Arnold approach. Nonetheless, a dramatically larger value of $K$ would signal the failure of our method to account for most of the dynamics relevant to ejection. Moreover, we verify in \cref{fig:lifetime-distribution} that our estimated ejection rate, interpreted as the rate of an exponential distribution, produces a good fit to the entire distribution of lifetimes extracted from simulations. As discussed in \cref{sec:capture-ejection-rate}, if the dynamics of ejection were not dominated by a single close encounter, we would expect a more general Erlang distribution rather than the simple exponential distribution seen here.

However, our main goal is to produce an estimate of the ejection time-scale that remains valid across a wide variety of systems. Thus, the real test of our result is the extent to which a fixed value of $K$ can be used to obtain an order-of-magnitude estimate of the ejection rate not only in the sun--Jupiter system, but in binaries with different mass ratios and semimajor axes. Indeed, even in the sun--Jupiter system, a single value of $K$ must be sufficient to predict the ejection rate for objects captured with many values of $v_\infty$.

\begin{figure}
    \includegraphics[width=\columnwidth]{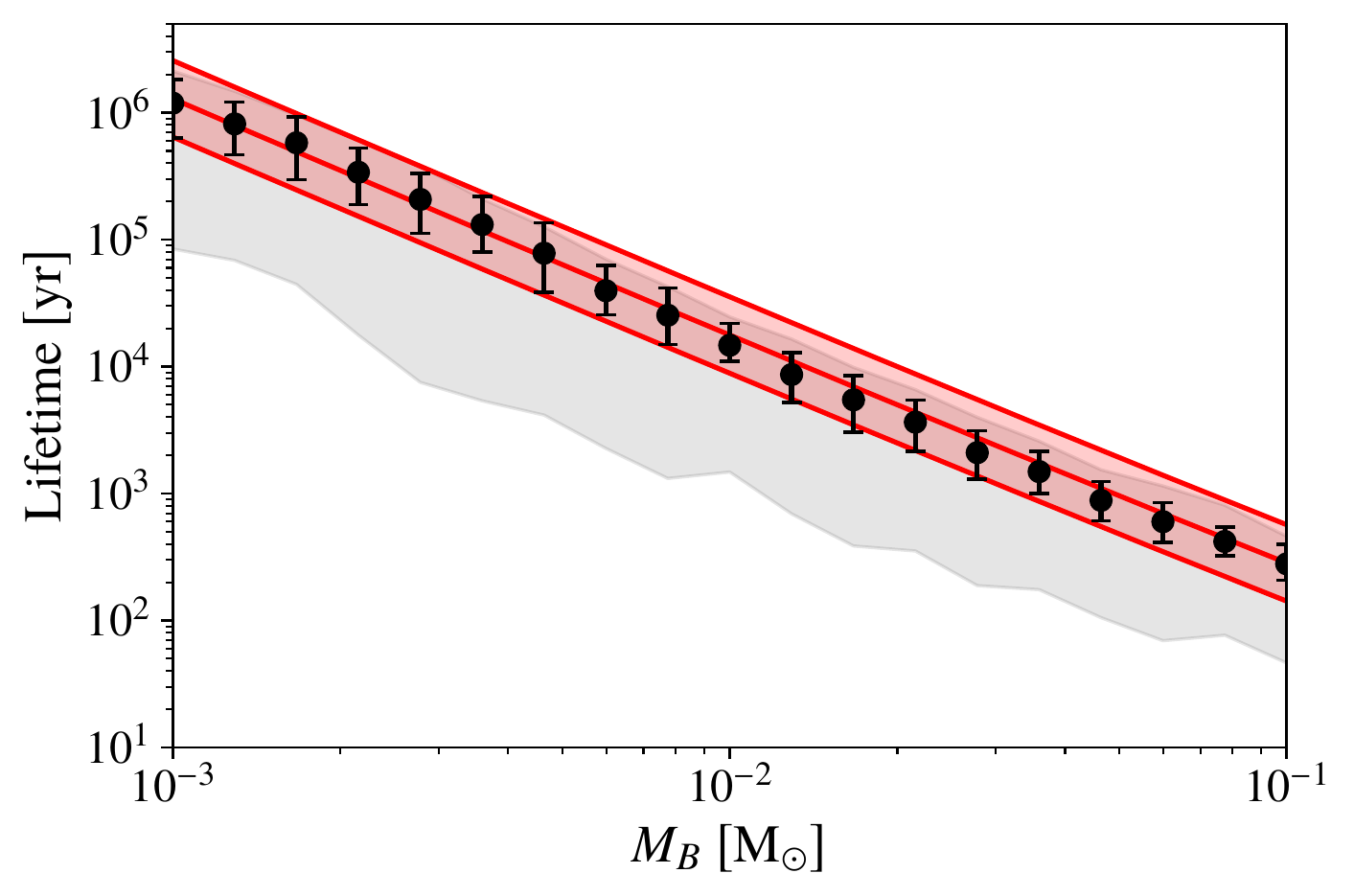}\\
    \includegraphics[width=\columnwidth]{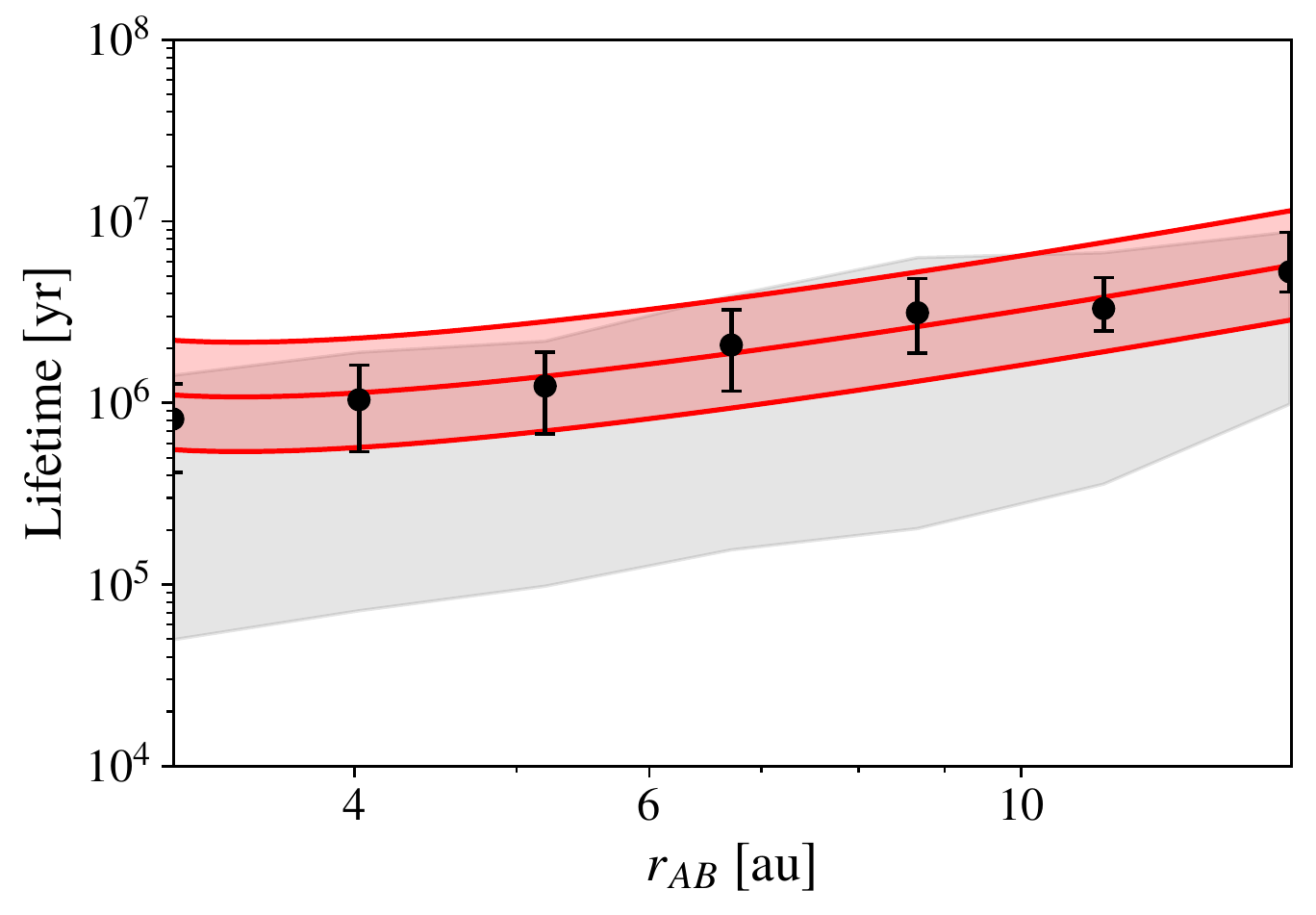}\\
    \includegraphics[width=\columnwidth]{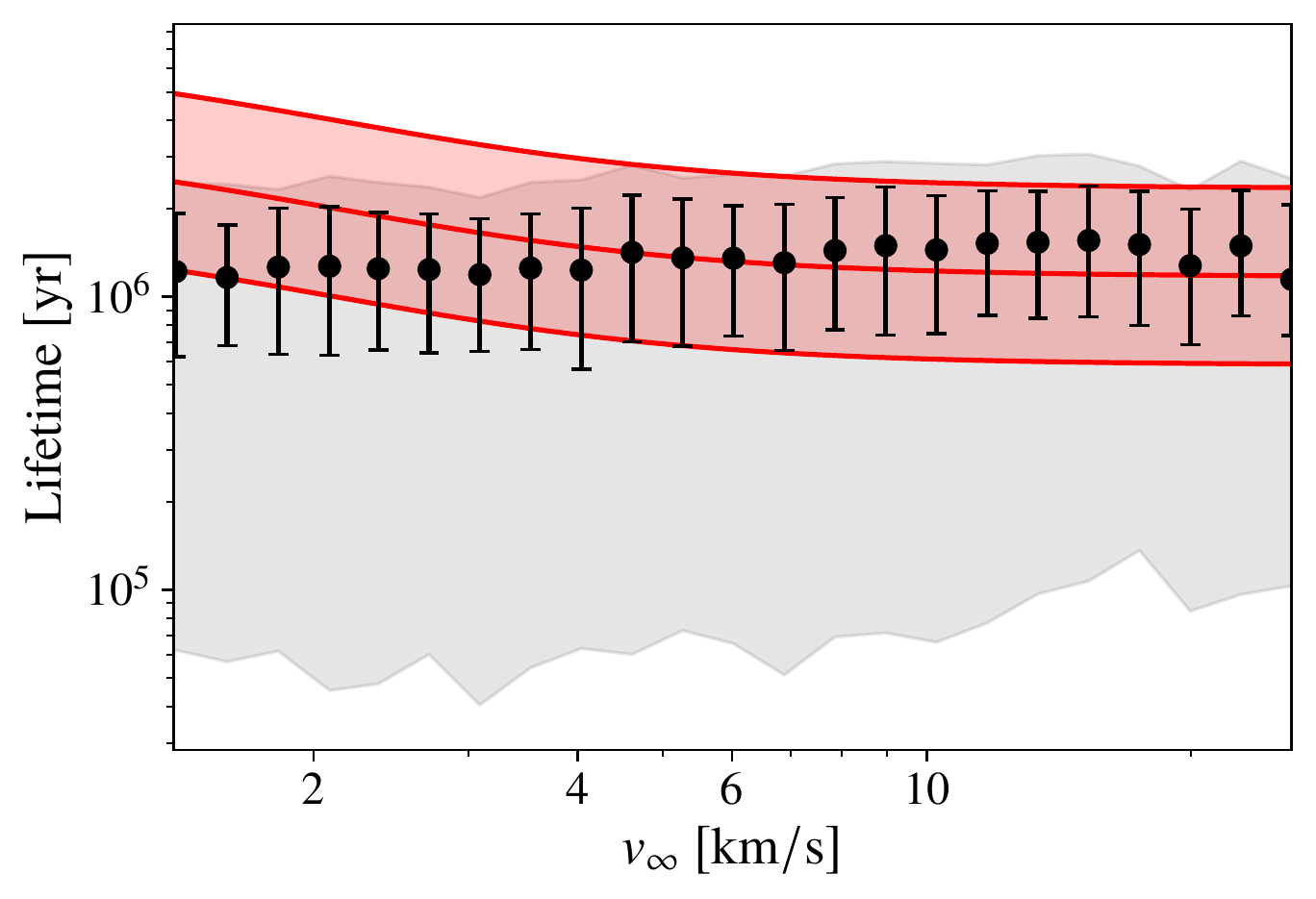}
    \caption{Mean ejection time-scale for captured objects as predicted by \cref{eq:ejection-rate} (red) and in an ensemble of numerical simulations (black). Error bars show $\pm1\sigma$ bootstrap confidence intervals. The gray regions show $\pm1\sigma$ quantiles for the lifetime distribution at each point. In each panel, one parameter is varied with respect to the base configuration, consisting of the sun--Jupiter system with $v_\infty=\SI{20}{\kilo\meter\per\second}$. The top panel varies the companion mass $M_B$, the middle panel varies the binary radius $r_{AB}$, and the bottom panel varies the initial speed $v_\infty$ of object $C$ prior to capture. The analytical prediction is shown for three values of $K$: $50$, $25$, and $12.5$ from bottom to top. Each black point shows the mean time to ejection after capture in an ensemble of simulations with randomized initial configurations. Note that at very small values of $v_\infty$, and potentially $r_{AB}$, our prediction becomes unreliable (see text).}
    \label{fig:capture-lifetimes}
\end{figure}

We thus vary these parameters and compare the outcomes of numerical simulations with the analytical prediction, with the results shown in \cref{fig:capture-lifetimes}. Some of the behavior in these results is easy to understand: in particular, the $M_B$ dependence can be estimated by the impact on the ejection cross section. Na\"ively, increasing the mass $M_B$ of the companion increases the ejection cross section as $\sej\sim M_B^2$, comparably decreasing the ejection time-scale. On the other hand, the dependence of the analytical estimate on $r_{AB}$ and $v_\infty$ is much more complicated. Note that even the $M_B$ dependence is not as straightforward as our heuristic argument would suggest, because the orbital parameter distribution of captured objects also has non-trivial $M_B$ dependence. Thus, even for this case, we must rely on the numerical results to benchmark the analytical calculation. \Cref{fig:capture-lifetimes} shows that \cref{eq:ejection-rate} provides an excellent order-of-magnitude estimate of the ejection time-scale, generally lying within a factor of 2 of the numerical mean.

Finally, we note that for some parameter values, the lifetime distribution is sensitive to the approximations that we make in deriving the orbital parameter distributions. In particular, for small values of $v_\infty$, our formalism can fail to accurately predict the distribution of semimajor axes after capture, resulting in disagreement between the analytical result and simulation outputs (see \cref{fig:capture-lifetimes}, bottom panel). This is to be expected due to tidal forces. Our approach assumes that the capture is driven by a close encounter, i.e., $\min r_{BC}\lesssim\rclose(\epsilon)$ (see \cref{sec:capture-cross-section}). But for small values of $v_\infty$, the capture cross section becomes large, and in particular, it is possible that $\sqrt{\scap}\gtrsim\rclose(\epsilon)$. In this case, the close-encounter condition is not satisfied for all impact parameters leading to capture, and our estimate of the orbital parameter distributions breaks down. A similar condition is produced by taking small values of $r_{AB}$, which causes $\rclose(\epsilon)$ to shrink.

\section{Discussion and conclusions}
\label{sec:discussion}
In the preceding sections, we have developed a simple formalism for identifying the parameters of close encounters that lead to capture. We have used this technology to consistently study the long-term behavior of such captured orbits. We now discuss the utility of our results and compare with other approaches in the literature.

There are several existing approaches to computing averaged capture rates and ejection rates, discussed at length by \citet{Valtonen:2005}. Our approach is distinct in several ways: First, our results accurately account for the full range of dependence on initial parameters, including arbitrary inclination, binary phase, and approach velocity $v_\infty$ of the third body. Secondly, we obtain a geometric classification of close encounters resulting in captures, and thirdly, we have applied this geometric result to develop a novel treatment of the ejection time-scale for captured objects based on \"Opik--Arnold theory. A valuable feature of our formalism is that it is easy to fix certain parameters and perform a detailed study of the distribution of captured orbits, with valuable applications to characterising the dynamics of captured objects in general binary systems. While we validate our analytical results against a set of numerical simulations, our formalism applies to a large class of binaries and can be readily deployed for analytical estimates in general extrasolar systems.

We have also applied our estimates of the capture and ejection rates to study the properties of the equilibrium population of captured objects. Our fiducial example concerns the capture of dark matter particles, and we find order-of-magnitude agreement with intricate numerical simulations. Note that although this procedure can produce a rough estimate, there are several conditions under which it might \emph{underestimate} the typical number density of captured objects. First, since the distribution of ejection time-scales has a long tail, some captures are much longer lived than most. Thus, for instance, $\langle\widetilde N\rangle\ll 1$ does not guarantee that there will not be even one object bound at any given time. It is also possible that our fairly rudimentary simulations fail to resolve the extent of this tail. Secondly, a set of especially long-lived outliers may be generated due to interactions with other bodies in the system, e.g., the smaller planets in the solar system, which we omit in our simulations.

In fact, our approach completely neglects interactions between the third body and any \emph{fourth} body not involved in the capture process, which is not always a valid approximation. For example, on occasion, resonances with planetary orbits may push the captured object into a much more stable orbit \citep{Lundberg:2004dn,Peter:2009mm}. This effect is irrelevant for capture by a binary system on its own, but is certainly relevant in the solar system, or in extrasolar systems with several close bodies. Our approach is also inappropriate in cases where the third body has mass comparable to the objects in the binary, where exchange or binary destruction are realizable.

However, a significant motivation for this work is the study of compact object binaries that may be used as probes of new physics. Many such systems are relatively simple, dynamically speaking, and can indeed be modelled spectacularly well. For instance, black hole binaries and their mergers are themselves detected through the use of well-understood template signals \citep{Balasubramanian:1995bm}, and pulsar binaries can be modelled so precisely that their dynamics probe general relativity, low-frequency gravitational waves, and compact object flybys \citep[see e.g.][]{Backer:1986wa,Stairs:2003eg,Cordes:2005gp,Hobbs:2009yy,Dror:2019twh,Ramani:2020hdo}. Our interest lies exactly in the non-trivial dynamics of capture by such binary systems. It is known that pulsar binaries in particular can sensitively probe the presence of captured compact objects \citep{1992ApJ...387L..69T,1993ApJ...407..266M}. With a detailed understanding of the statistics of capture, it is in principle possible to use capture as a mechanism to constrain the population of compact objects such as primordial black holes \citep[see e.g.][]{Carr:2003bj,Carr:2016drx}, ultracompact minihaloes \citep{Ricotti:2009bs}, and other exotic objects. We will explore this possibility in detail in a subsequent study.

\section*{Acknowledgements}
BVL and SP are partly supported  by the U.S. Department of Energy grant number DE-SC0010107. We gratefully acknowledge valuable conversations with Ruth Murray-Clay, Abraham Loeb, Smadar Naoz, and Hagai Perets. Simulations in this paper made use of the \textsc{rebound} code, which is freely available at \href{http://github.com/hannorein/rebound}{\textcolor{blue}{http://github.com/hannorein/rebound}}.

\section*{Data Availability}
The data underlying this article will be shared on reasonable request to the corresponding author.

\bibliographystyle{mnras}
\bibliography{main}

\bsp
\label{lastpage}
\end{document}